\documentclass[twocolumn,
 superscriptaddress,
 amsmath,amssymb,
 aps,
 floatfix
]{revtex4-2}

\usepackage{amsmath}
\usepackage{amssymb}
\usepackage{amsthm}
\usepackage{bm}
\usepackage{graphicx}
\usepackage{hyperref}
\usepackage[capitalise]{cleveref}

\begin{document}

\title{Higher Order Dynamics in the Replicator Equation Produce a Limit Cycle in Rock-Paper-Scissors}

\author{Christopher Griffin}
\email{griffinch@psu.edu}
\affiliation{
	Applied Research Laboratory,
    The Pennsylvania State University,
    University Park, PA 16802
    }

\author{Rongling Wu}
\email{ronglingwu@mail.tsinghua.edu.cn}
\affiliation{ Beijing Institute of Mathematical Sciences and Applications, Beijing 101408, China}
\affiliation{Yau Mathematical Sciences Center, Tsinghua University, Beijing 100084, China}

\begin{abstract} Recent work has shown that pairwise interactions may not be sufficient to fully model ecological dynamics in the wild. In this letter, we consider a replicator dynamic that takes both pairwise and triadic interactions into consideration using a rank-three tensor. We study {these} new nonlinear dynamics using a generalized rock-paper-scissors game whose dynamics are well understood in the {standard} replicator sense. We show that the addition of higher-order dynamics leads to the creation of a subcritical Hopf bifurcation and consequently an unstable limit cycle. It is known that this kind of behaviour cannot occur in the pairwise replicator in any three strategy games, showing the effect higher-order interactions can have on the resulting dynamics of the system. We numerically characterize parameter regimes in which limit cycles exist and discuss possible ways to generalize this approach to studying higher-order interactions.
\end{abstract}

\maketitle

\maketitle

\section{Introduction}
Pairwise interactions are frequently assumed in constructing dynamical systems models of ecological systems \cite{M72,P81,RY02,KDGL22}. This is a foundational assumption of classical evolutionary game theory, in which the replicator dynamic is built from a {game matrix} \cite{W97,HS98,HS03}. In this case, pairwise interactions of players define the fitness function that governs the dynamics. This simplifying assumption is violated by the intrinsic existence of higher-order interactions (HOIs), for which there is growing evidence \cite{LBAA17,GBMA17,BKK16, MFFS19,SATA21,KKPL22,GLL22,BABB21,BCIL20,LRS19,SFF22}. A higher-order interaction occurs when three or more species act together as a subgroup to shape community behaviour\cite{P81,LBAA17,GBMA17,BKK16, MFFS19,SATA21,MS17, MK19,DTS22}. In the case of random interactions in ecological communities, the occurrence of HOIs can alter the established relationship between diversity and stability \cite{BKK16}, leading to new evolutionary trajectories.

In this letter, we show how to modify the standard game matrix replicator dynamic by defining fitness in terms of both a game matrix (for pairwise interactions) and a game tensor (for triadic interactions). Early work exhibiting limit cycles and higher-order interactions is by Hofbauer, Schuster and Sigmund \cite{HSS82}. Additionally, this work is related to the prior work of Gokhale and Traulsen \cite{GT10} who studied evolutionary games with multiple (more than two) strategies and multiple players {using a game tensor}. In this work, they study the maximum number of mixed strategy equilibria that may emerge in the {resulting} replicator dynamic. More recently, Zhang et al. \cite{ZPZW22} study multiplayer evolutionary games in the context of asymmetric payoffs, which we do not consider. {Additionally, Peixe and Rodrigues \cite{PR22a} study strange attractors and super-critical Hopf bifurcations in \textit{polymatrix replicators}, modelling inter and intra group interactions in a multi-group population. Polymatrix games are also considered in \cite{AD15,PG16}.} {In contrast,} we study the resulting dynamics on a generalized rock-paper-scissors game (RPS) \cite{W97} with higher-order interactions in a single population. We show that the resulting dynamics arising from triadic interactions are fundamentally different from those dynamics arising from RPS in the {standard} replicator equation with only pairwise interactions by showing that the HOIs lead to the emergence of a limit cycle. 

Rock-paper-scissors (and its {generalizations}) has been studied in multiple different contexts \cite{ML75,M10,PR19,HDL13, SMR14,SMR13,SMJS14,RMF08,RMF07,PR17,BMDR17,HMT10,KT21,GSB22,PR22,MMP19,AOT21,P19,MASB19,MBTT22,I87,B88,VS93,G21, PG22} and there are at least two schools of partially compatible dynamics. Postlewaite and Rucklidge \cite{PR17,PR19,PR22,SMJS14,SMR13,SMR14} have extensively studied a dynamical systems model of RPS in both spatial and aspatial cases. Their dynamics are distinct \cite{GMd21} from the dynamics arising from the {standard} replicator equation, as given in \cite{W97,HS03,HS98}. We do not consider their dynamics, but instead focus on {those arising from} the replicator equation. In particular, Zeeman \cite{Z80} showed that RPS dynamics under the {standard} replicator exhibit a degenerate Hopf bifurcation and cannot produce limit cycles. More generally, Zeeman showed that no three strategy game can produce a limit cycle under the {standard} replicator dynamics. However, since every dynamical system arising from the {standard} replicator equation is diffeomorphic to a generalized Lotka-Volterra system, limit cycles and chaos may emerge for games with more than three strategies. 

The main results of this letter are: (i) We {propose a method for modelling} triadic interactions using a simple {rank-three} tensor. (ii) We show (numerically) that in generalized RPS with HOIs a subcritical Hopf bifurcation occurs, and a limit cycle emerges for appropriate parameter choices. This behaviour must be caused by the HOIs, since such dynamics cannot emerge with only pairwise interactions \cite{Z80}. (iii) We use a statistical analysis to construct a two-dimensional bifurcation surface, showing parameter regions where the (unique) interior fixed point is stable and admits a limit cycle, is stable with no limit cycle, and is unstable.

\section{General Model}
Let $\Delta_{n-1}$ be the $n-1$ dimensional unit simplex embedded in $\mathbb{R}^n$ composed of vectors $\mathbf{u} = \langle{u_1,\dots,u_n}\rangle$ so that $u_1 + \cdots + u_n = 1$ and $u_i \geq 0$ for all $i\in\{1,\dots,n\}$. {In a biological context, $u_i$ is the proportion of species $i$ when considered as part of the total biomass to be modelled.} 

{Let $\mathbf{A} \in \mathbb{R}^{n \times n}$ be a payoff matrix and assume that species $i$ receives an expected payoff
\begin{equation*}
f_i(\mathbf{u}) = \mathbf{e}_i^T\mathbf{A}\mathbf{u},
\end{equation*}
as a result of both inter and intra species interactions. Here $\mathbf{e}_i$ is the $i^\text{th}$ standard basis vector in Euclidean space. The \textit{standard replicator equation} that arises from this fitness function is given by
\begin{equation}
\dot{u}_i = u_i\left[f(\mathbf{u}) - \bar{f}(\mathbf{u})\right] = u_i\left(\mathbf{e}_i^T\mathbf{A}\mathbf{u} - \mathbf{u}^T\mathbf{A}\mathbf{u}\right),
\label{eqn:StandardReplicator}
\end{equation}
where $\bar{f}(\mathbf{u}) = \mathbf{u}^T\mathbf{A}\mathbf{u}$.} \cref{eqn:StandardReplicator} assumes simple binary interactions among species, with the payoff to species $i$ resulting from an interaction between species $i$ and species $j$ given by $\mathbf{A}_{ij}$.

{To model higher-order interactions, redefine} $f_i(\mathbf{u})$ to be,
\begin{equation}
f_i(\mathbf{u}) = \mathbf{e}_i^T\mathbf{A}\mathbf{u} + \mathbf{u}^T\mathbf{B}_i\mathbf{u},
\label{eqn:f-HigherOrder}
\end{equation}
where $\mathbf{B}_i$ is a quadratic form (matrix) that takes two copies of the population proportion vector $\mathbf{u}$ and returns a payoff to species $i$ that occurs when one member of  {species} $i$ meets two members of the population {(at random)}. In general, we could think of $\mathbf{B}_i$ as being a slice of a $(0,3)$ tensor $B:\Delta_{n-1}\times\Delta_{n-1}\times\Delta_{n-1} \to \mathbb{R}$. The replicator equation still has form,
\begin{equation}
\dot{u}_i = u_i\left[f(\mathbf{u}) - \bar{f}(\mathbf{u})\right], 
\label{eqn:GeneralReplicator}
\end{equation}
with
{\begin{multline}
\bar{f} = \sum_{i = 1}^n u_i f_i(\mathbf{u}) = \sum_{i=1}^n u_i \left( \mathbf{e}_i^T\mathbf{A}\mathbf{u} + \mathbf{u}^T\mathbf{B}_i\mathbf{u} \right) = \\
\mathbf{u}^T\mathbf{A}\mathbf{u} + \sum_{i=1}^n u_i\mathbf{u}^T\mathbf{B}_i\mathbf{u}.
\label{eqn:fbar}
\end{multline}
}

We now propose a biologically inspired approach to defining $\mathbf{B}_i$. In what follows, we will define 
\begin{equation*}
B_{i,j,k} = p_{ijk}\mathbf{A}_{ij} + q_{ijk}\mathbf{A}_{ik},
\end{equation*}
for constants of proportionality $p_{ijk}$ and $q_{ijk}$. That is, we assume that the three-way payoff is composed of payoffs from binary interactions that are scaled to model the effects of the more complex interactions. In our analysis of generalized rock-paper-scissors, we choose $p_{ijk}$ and $q_{ijk}$ so that the Nash equilibrium of RPS remains a fixed point. It is left as a question for future work whether there are sufficient conditions on the tensor $B$ that ensure the Nash equilibria of the game matrix $\mathbf{A}$ are preserved as fixed points in three-way dynamics.

Before proceeding to the analysis of RPS, we note that \cref{eqn:f-HigherOrder} could be generalized to include $n > 3 $-way interactions by using higher rank tensors \cite{GT10}. However, it is unlikely that such interactions are biologically meaningful. Statistical tests for these kinds of interactions are discussed in \cite{FLHJ23}. 

\section{Higher Order Rock-Paper-Scissors}
The remainder of this paper is dedicated to showing that higher-order-interactions in a generalized rock-paper-scissors game produce dynamics not seen when only pairwise interactions are modelled. {Fix the parameterized payoff matrix},
\begin{equation}
\mathbf{A} = 
\begin{bmatrix}
0 & -b-1 & a+1 \\
a+1 & 0 & -b-1 \\
-b-1 & a+1 & 0 
\end{bmatrix},
\label{eqn:GeneralizedRPS}
\end{equation}
where we assume $a,b\geq 0$. {When $a = b = 0$, this is the standard RPS matrix found in (e.g.) \cite{GJW20,G21,GMd21}.} This matrix will govern the payoff from simple pairwise interactions. We now define the tensor $B$ using its slices, so that
\begin{align}
\mathbf{B}_1 &= 
\begin{bmatrix}
0 & \frac{1}{2} (-b-1) & \frac{1}{2} (a+1) \beta  \\
\frac{1}{2} (-b-1) & -b-1 & 0 \\
\frac{1}{2} (a+1) \beta  & 0 & (a+1) \alpha 
\end{bmatrix}\label{eqn:B1}\\
\mathbf{B}_2 &= 
\begin{bmatrix}
(a+1) \alpha  & \frac{1}{2} (a+1) \beta  & 0 \\
\frac{1}{2} (a+1) \beta  & 0 & \frac{1}{2} (-b-1) \\
0 & \frac{1}{2} (-b-1) & -b-1 
\end{bmatrix}\label{eqn:B2}\\
\mathbf{B}_3 &= 
\begin{bmatrix}
-b-1 & 0 & \frac{1}{2} (-b-1) \\
 0 & (a+1) \alpha  & \frac{1}{2} (a+1) \beta  \\
 \frac{1}{2} (-b-1) & \frac{1}{2} (a+1) \beta  & 0 
\end{bmatrix}\label{eqn:B3}.
\end{align}
Here we assume $\alpha \in (1,2]$ and $\beta \in \left(0, {1}\right]$ for {simplicity}. The reasoning behind defining $B$ in this way is justified by considering the payoff associated to rock (index 1). If a rock plays against two other rocks, it receives no payoff -- as expected from \cref{eqn:GeneralizedRPS}. If it meets two papers {(index 2)}, then it receives the same net negative payoff $-b-1$ as if it met one paper. {If it meets both a rock and paper, then its net negative payoff is cut in half, since the paper plays against two rocks.} On the other hand, if it meets two scissors {(index 3)}, then its payoff increases by a factor of $\alpha$. If a rock meets a rock and scissors, then the rocks split the payoff and each receives a payoff decreased by a factor $\beta$, caused by competition between the two rocks. Finally, if a rock meets both a paper and scissors, then mutual destruction leads to no net payoff for any player. The same logic {\textit{mutatis mutandis}} is used for all other players. 

\cref{eqn:B1,eqn:B2,eqn:B3} enforce symmetry in each slice and preserve a generalized cyclic property. Note that {$\mathbf{B}_2$} can be recovered from {$\mathbf{B}_1$} by rotating the rows and columns of $\mathbf{B}_1$ down and to the right (just as row $i+1$ can be obtained from row $i$ in $\mathbf{A}$ by rotating to the right). Thus, $B$ is a circulant (or Toeplitz) tensor. 

It is straightforward to see that {the general replicator dynamics given by \cref{eqn:f-HigherOrder,eqn:GeneralReplicator,eqn:fbar} using the matrix and tensor defined in \cref{eqn:GeneralizedRPS,eqn:B1,eqn:B2,eqn:B3} has $\bar{\mathbf{u}}_i =\mathbf{e}_i$ ($i\in\{1,2,3\}$) as fixed points. These correspond to single species populations or pure strategies. A fourth equilibrium is the interior point $\bar{\mathbf{u}} = \langle{\tfrac{1}{3},\tfrac{1}{3},\tfrac{1}{3}}\rangle$, corresponding to a perfectly  mixed population or the Nash equilibrium of RPS.} The fact that $\mathbf{A}$ is a circulant matrix and $B$ is a circulant tensor ensures that these are the only equilibria in the system {that occur} in $\Delta_2$ and $\bar{\mathbf{u}}$ is the only interior equilibrium. 

\subsection{Fixed Point Analysis}
The eigenvalues of the Jacobian matrix at any fixed point $\bar{\mathbf{u}}_i$ ($i\in\{1,2,3\}$) are
\begin{equation*}
\Lambda = \{0,-2 (b+1), (a+1) (\alpha +1)\}.
\end{equation*}
Based on the assumptions on the values of $a$, $b$ and $\alpha$, these are hyperbolic saddles {(see Theorem 3.3 of \cite{V06}). This is identical to the behaviour of the pure strategy equilibria in the case of ordinary rock-paper-scissors in the standard replicator dynamic \cite{HS03}.} 

{The Jacobian matrix at the interior fixed point $\bar{\mathbf{u}}$ has one real and two complex eigenvalues given by}
\begin{equation*}
\Lambda = \left\{\frac{r_1}{9}, \frac{r_2 \pm i\sqrt{3}q}{18}\right\},
\end{equation*}
where
\begin{align*}
r_1 &= 2 + 5b -\alpha -\beta -a (\alpha +\beta +3) \quad \text{and}\\
r_2 &= 1 + a (\beta - 2 \alpha- 3 ) + 4 b - 2 \alpha  + \beta
\end{align*}
and
\begin{equation*}
q = 9 + 6 b + 2 \alpha + \beta + a (3 + 2 \alpha + \beta).
\end{equation*}
The first (real) eigenvalue is extraneous, since the dynamics evolve on $\Delta_2$. 
Assuming $\alpha$, $\beta$ and $b$ are free, $r_2 = 0$ when
\begin{equation}
a = a^* = \frac{1 +4 b+\beta -2 \alpha }{3 + 2 \alpha -\beta}.
\label{eqn:astar}
\end{equation}
When $a = a^* + \epsilon$, then
\begin{equation*}
r_2 = \epsilon  (\beta -2 \alpha  -3).
\end{equation*}
If follows from our assumptions on $\alpha$ and $\beta$ that if $\epsilon > 0$, then  $a > a^*$ and $r_{2} < 0$ and {$\bar{\mathbf{u}}$} is attracting. Otherwise, {$\bar{\mathbf{u}}$} is repelling. For $\epsilon = 0$, the system has two pure imaginary eigenvalues, which satisfies the first requirement of Hopf's theorem (see \cite{GH13}, Page 152). Our assumptions that $\alpha \in (1,2]$ and $\beta \in \left(0, {1}\right]$ ensures that
\begin{equation*}
r_2'(\epsilon) = (\beta -2 \alpha  -3) \neq 0.
\end{equation*}
Thus, the real part of the eigenvalues must cross the imaginary axis with non-zero speed, satisfying the second criterion of Hopf's theorem. Thus, the system exhibits a Hopf bifurcation. 


\subsection{Numerical Illustration of a Subcritical Limit Cycle}
We can show numerically that an {unstable} limit cycle emerges for example parameters. For the remainder of this section, let $b = 0$. In our initial limit cycle construction, we assume $\alpha = 2$ and $\beta = \tfrac{1}{2}$. In this case, the interaction of (e.g.) a rock with two scissors doubles the payoff to {the rock}, while two rocks interacting with {one scissors} will quarter the payoff associated to the interaction. We set $\epsilon=\frac{1}{100}$, implying the interior fixed point will be stable.
\begin{figure}[htbp]
\centering
\includegraphics[width=0.8\columnwidth]{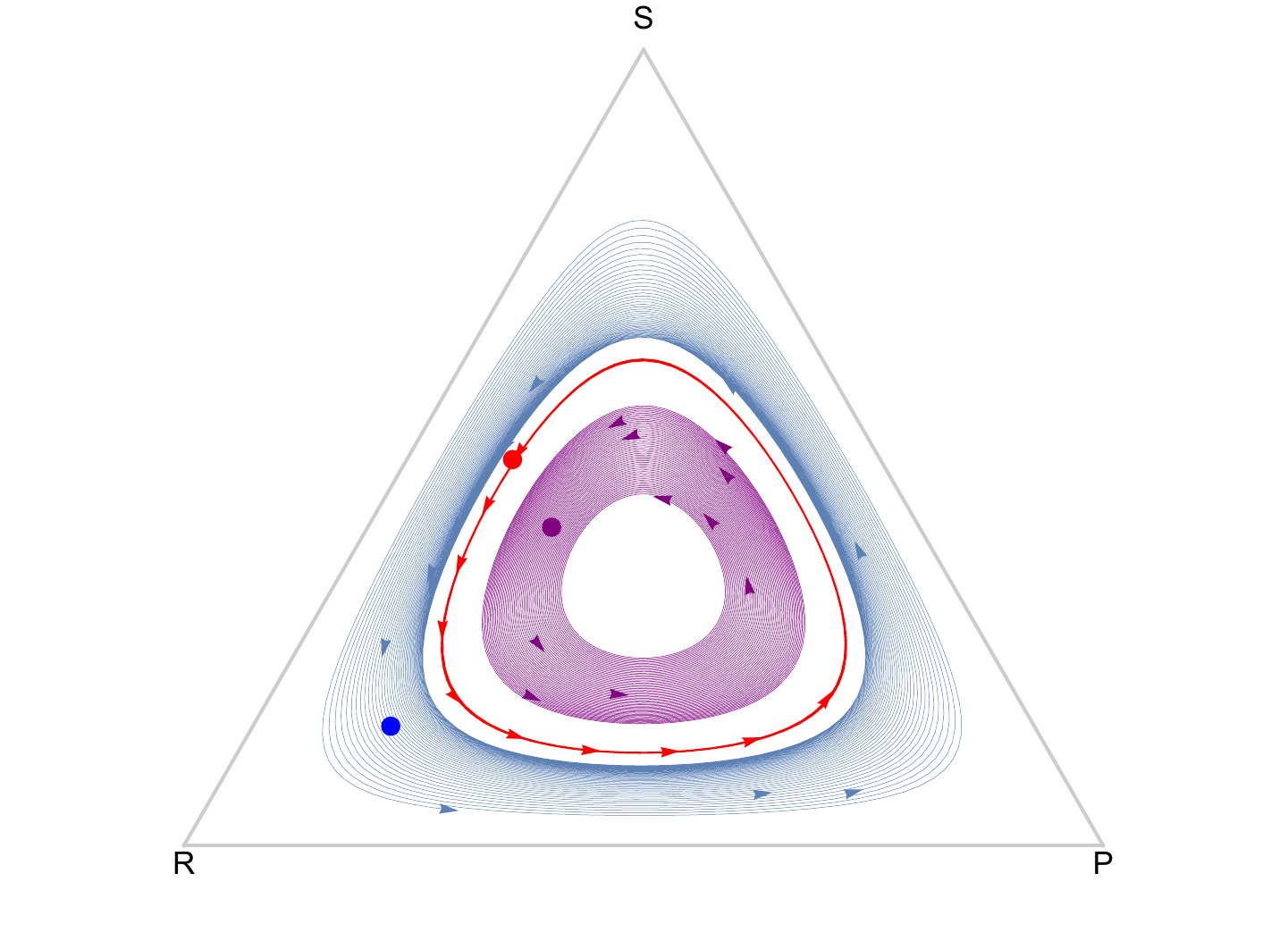}\\
\includegraphics[width=0.8\columnwidth]{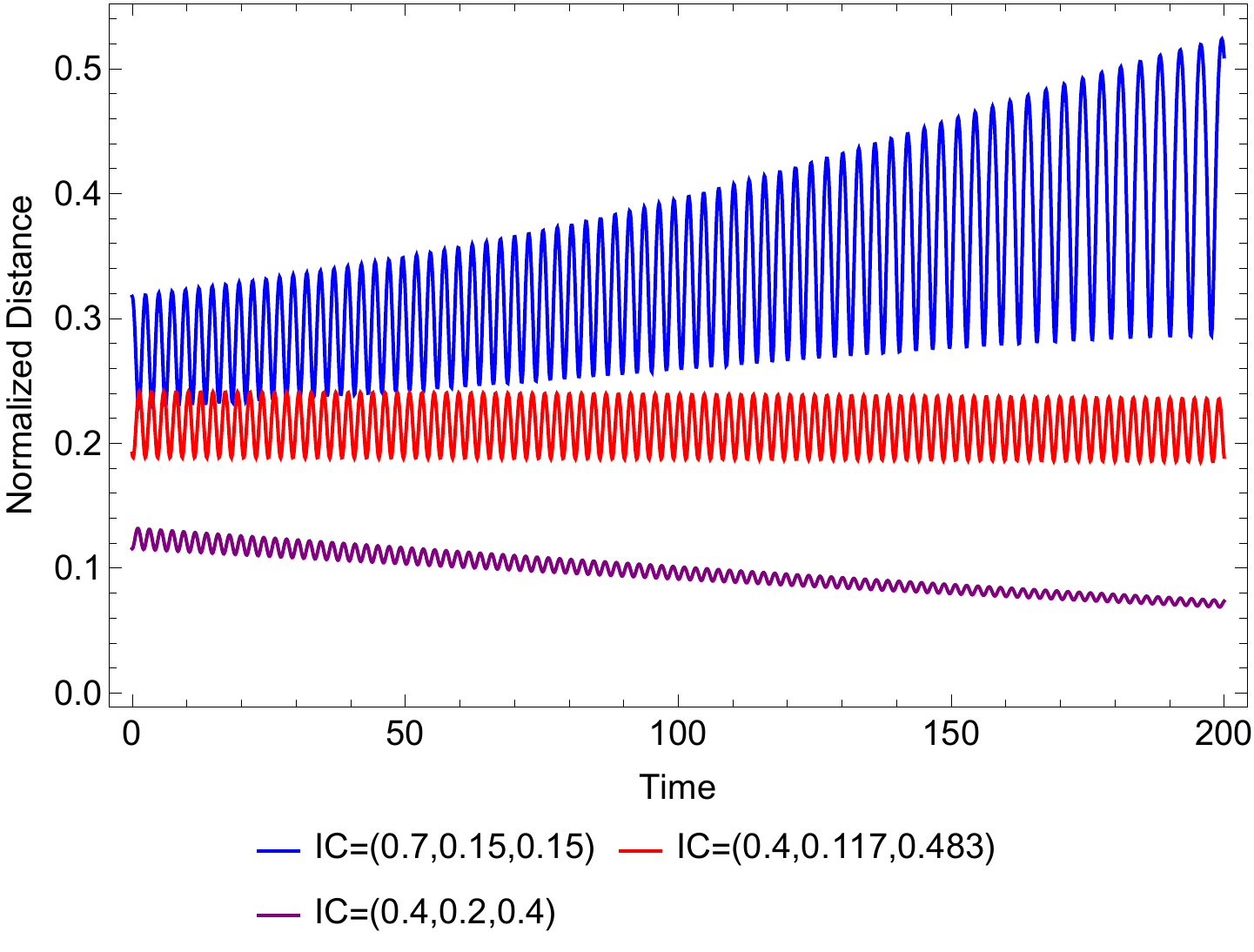} 
\caption{(Top) Example trajectories with a subcritical limit cycle are shown. (Bottom) The {distances} of the trajectories to the interior fixed point as a function of time. This proves numerically that there is a subcritical Hopf bifurcation in these dynamics via the Poincar\'{e}-Bendixson theorem.}
\label{fig:LimitCycle}
\end{figure}
In \cref{fig:LimitCycle} {(top)} we see an (approximated) limit cycle surrounding a stable interior fixed point. Outside the limit cycle, flow goes to the boundary. To complete the numeric proof, we apply the Poincar\'{e}-Bendixson theorem. In \cref{fig:LimitCycle} {(bottom)}, we compute the distance from the three trajectories to the interior fixed point after a ternary transform. This is a true representation of the distance seen in the trajectories in \cref{fig:LimitCycle} {(top)}. We can see that the distance from the (approximated) limit cycle {to the interior fixed point} oscillates around a constant mean. The trajectory outside the limit cycle increases its distance to the interior fixed point, while the trajectory inside the limit cycle decreases its distance to the interior fixed point as expected. Thus, the numerically identified limit cycle is {unstable, implying a subcritical Hopf bifurcation.}

\begin{figure}[htbp]
\centering
\includegraphics[width=0.8\columnwidth]{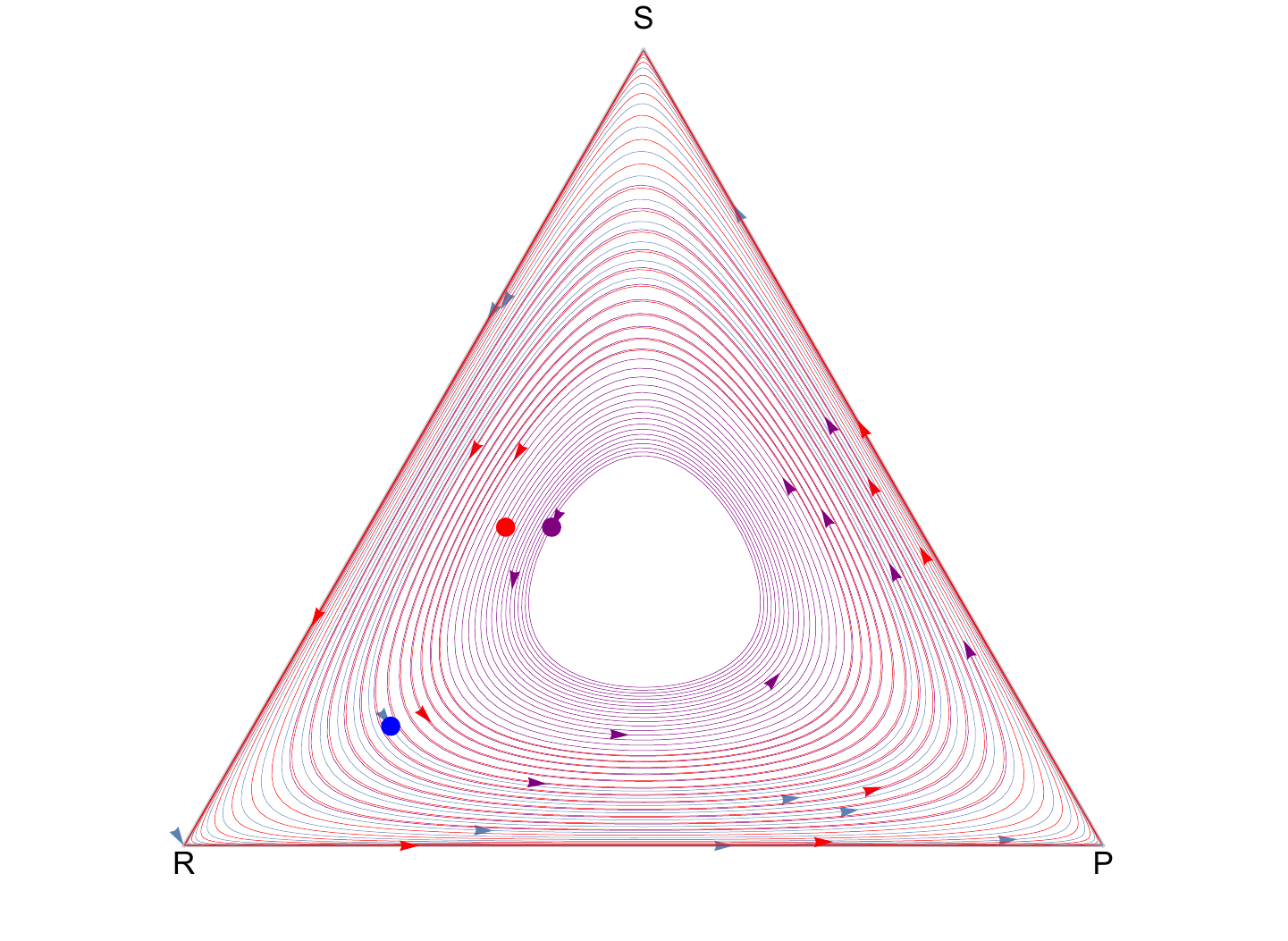}\\
\includegraphics[width=0.8\columnwidth]{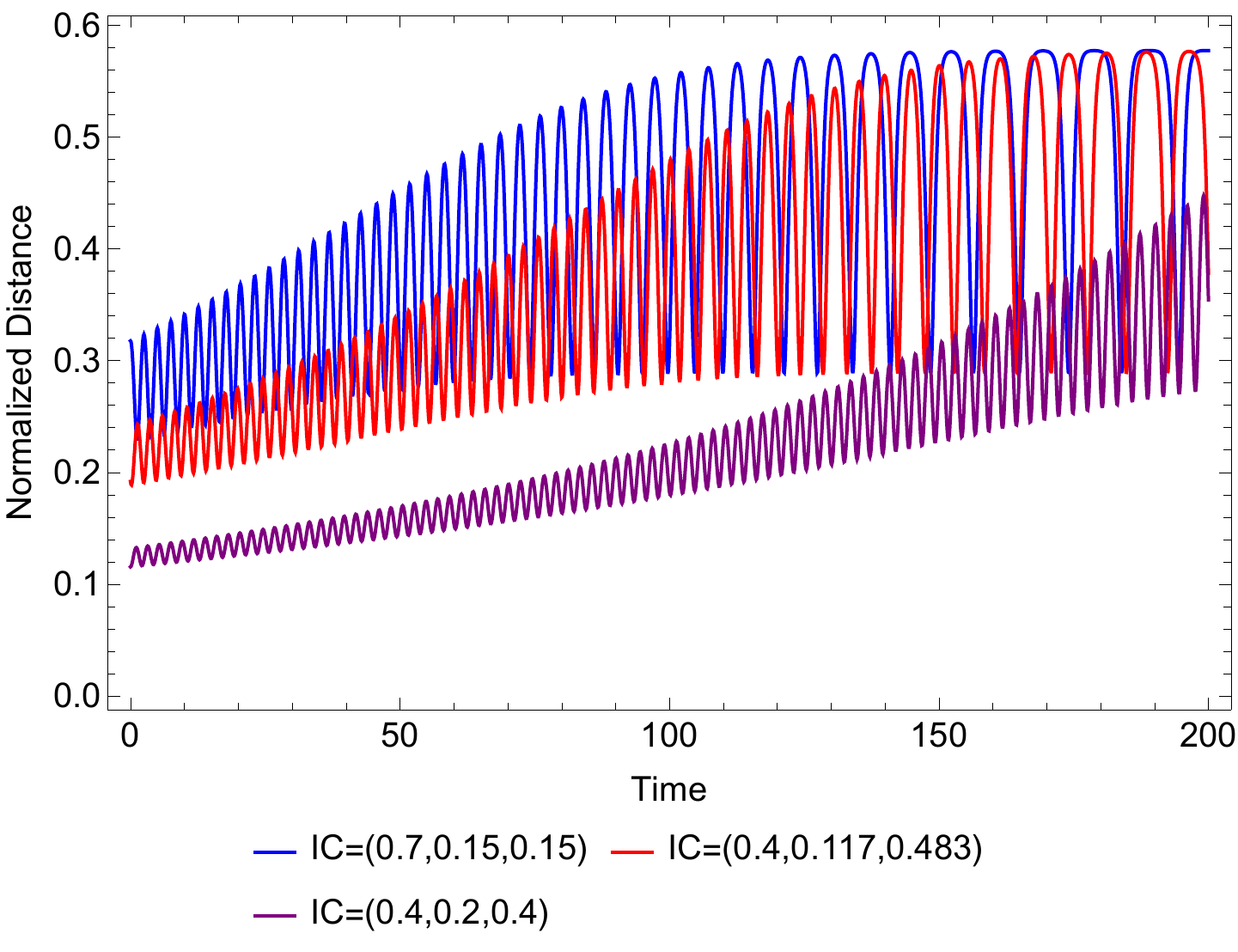}
\caption{(Top) {Setting} $\epsilon$ below zero destabilizes the interior fixed point and destroys the limit cycle (as expected). All trajectories approach the boundary. (Bottom) The distances of the trajectories to the interior fixed point increase as the trajectories approach the boundary of $\Delta_2$.}
\label{fig:Unstable}
\end{figure}

When $\epsilon < 0$, the interior fixed point becomes unstable (as expected) and the limit cycle vanishes as shown in \cref{fig:Unstable} {(top)}.
All trajectories approach the boundary, as confirmed numerically in \cref{fig:Unstable} {(bottom)}, which again shows the normalized distance from the trajectory to the interior fixed point. 

\begin{figure}[htbp]
\centering
\includegraphics[width=0.8\columnwidth]{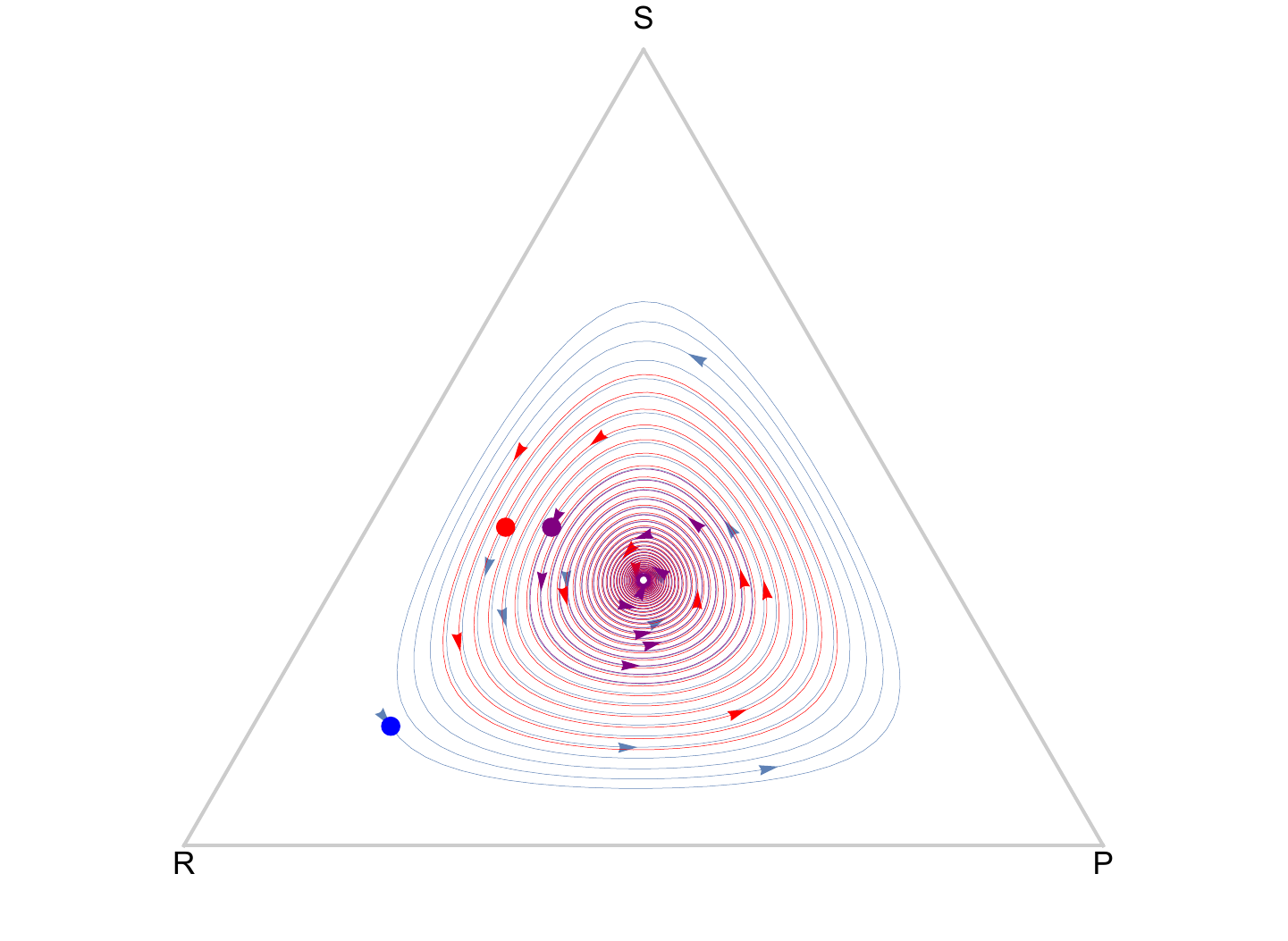}\\
\includegraphics[width=0.8\columnwidth]{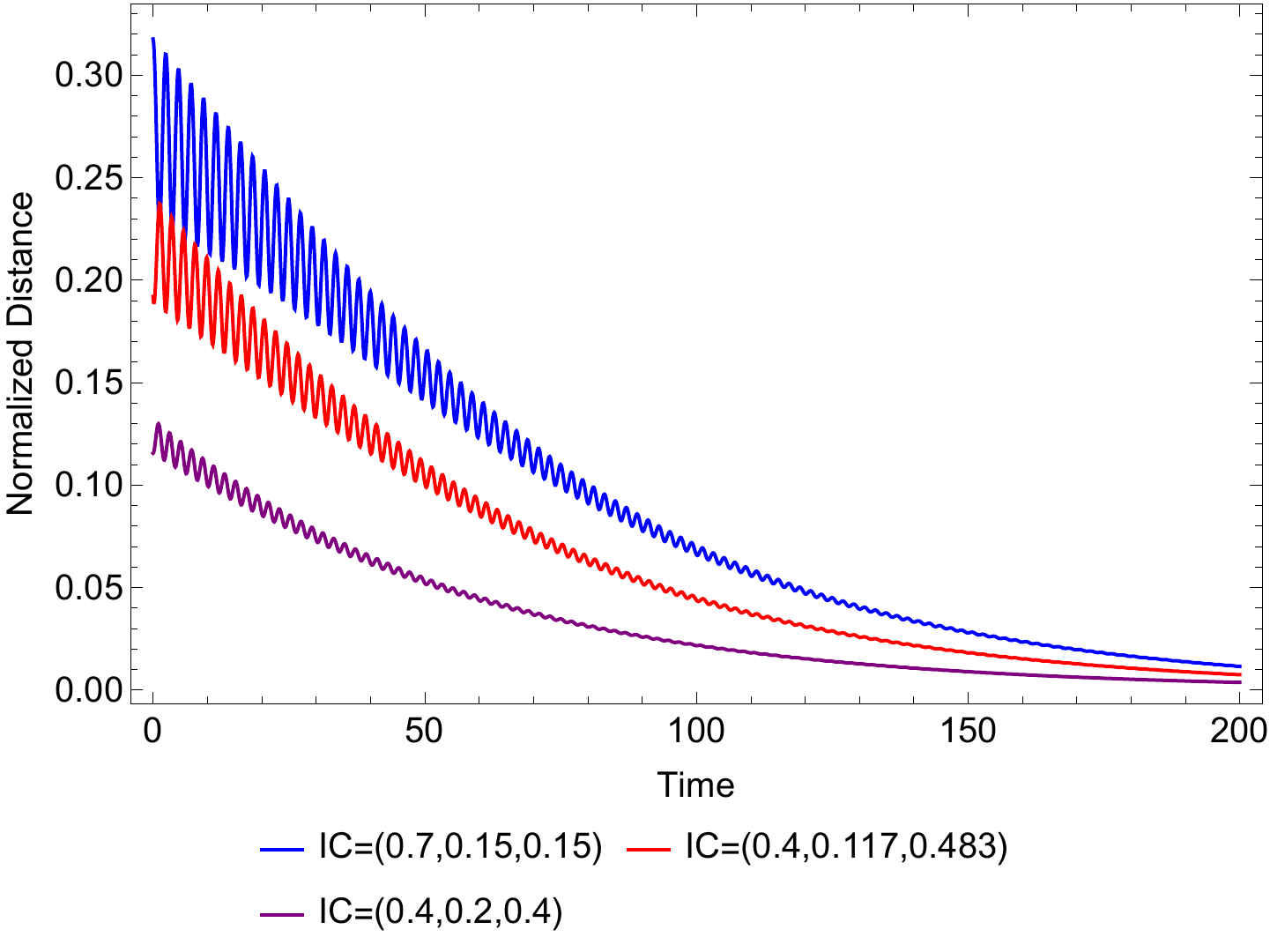}
\caption{(Top) If $\epsilon$ is increased beyond a critical value, the limit cycle is destroyed and all trajectories approach the interior fixed point. (Bottom) The distances of the trajectories to the interior fixed point decrease as expected.}
\label{fig:Stable}
\end{figure}

When $\epsilon$ is increased beyond a certain value, the limit cycle disappears while the interior fixed point remains stable. This is illustrated in \cref{fig:Stable} {(top)}. 
In \cref{fig:Stable} {(bottom)}, we show that the distance from any trajectory to the interior fixed point collapses to zero as time goes to infinity.

\begin{figure}[htbp]
\centering
\includegraphics[width=0.95\columnwidth]{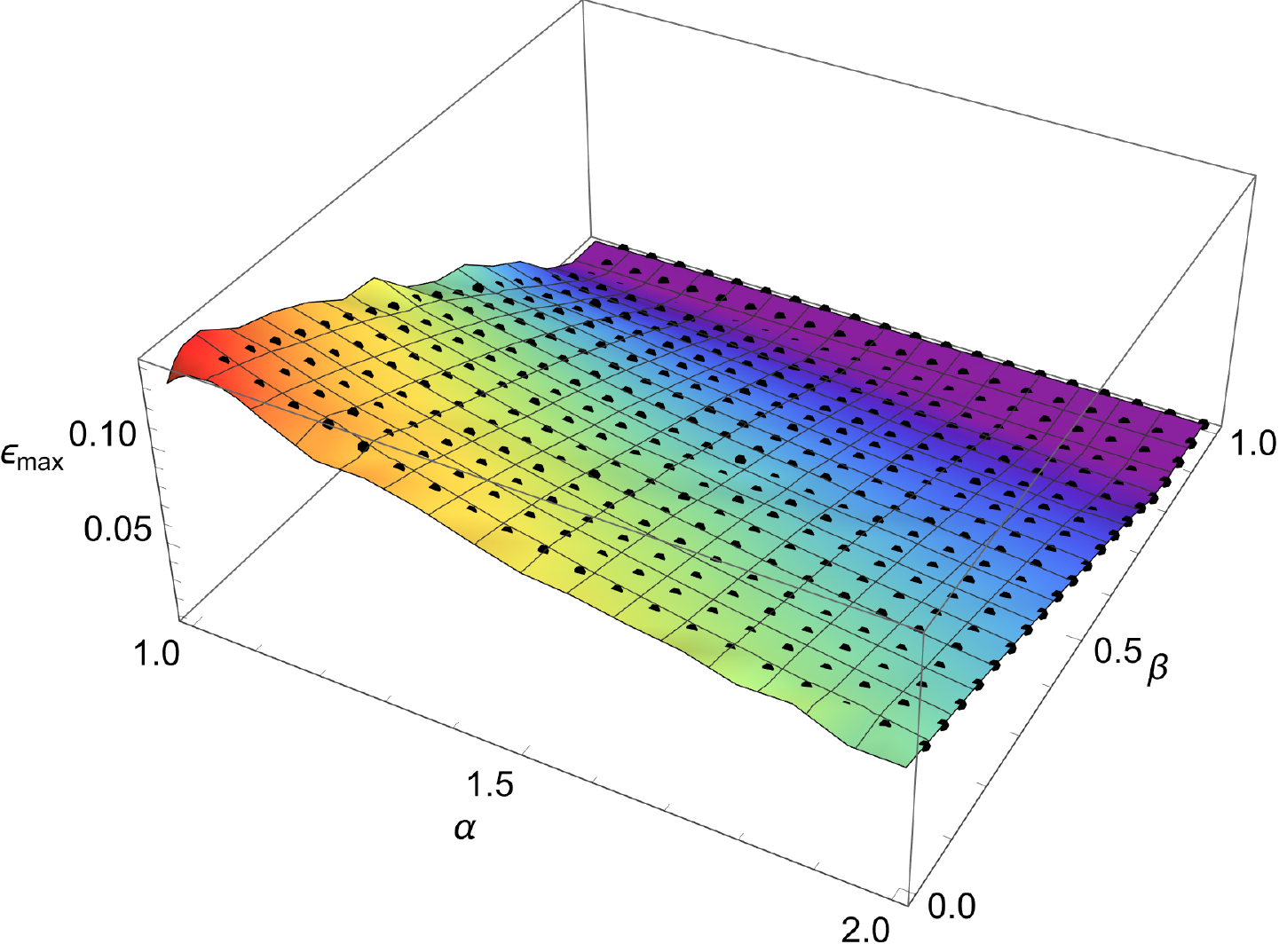}\\
\includegraphics[width=0.95\columnwidth]{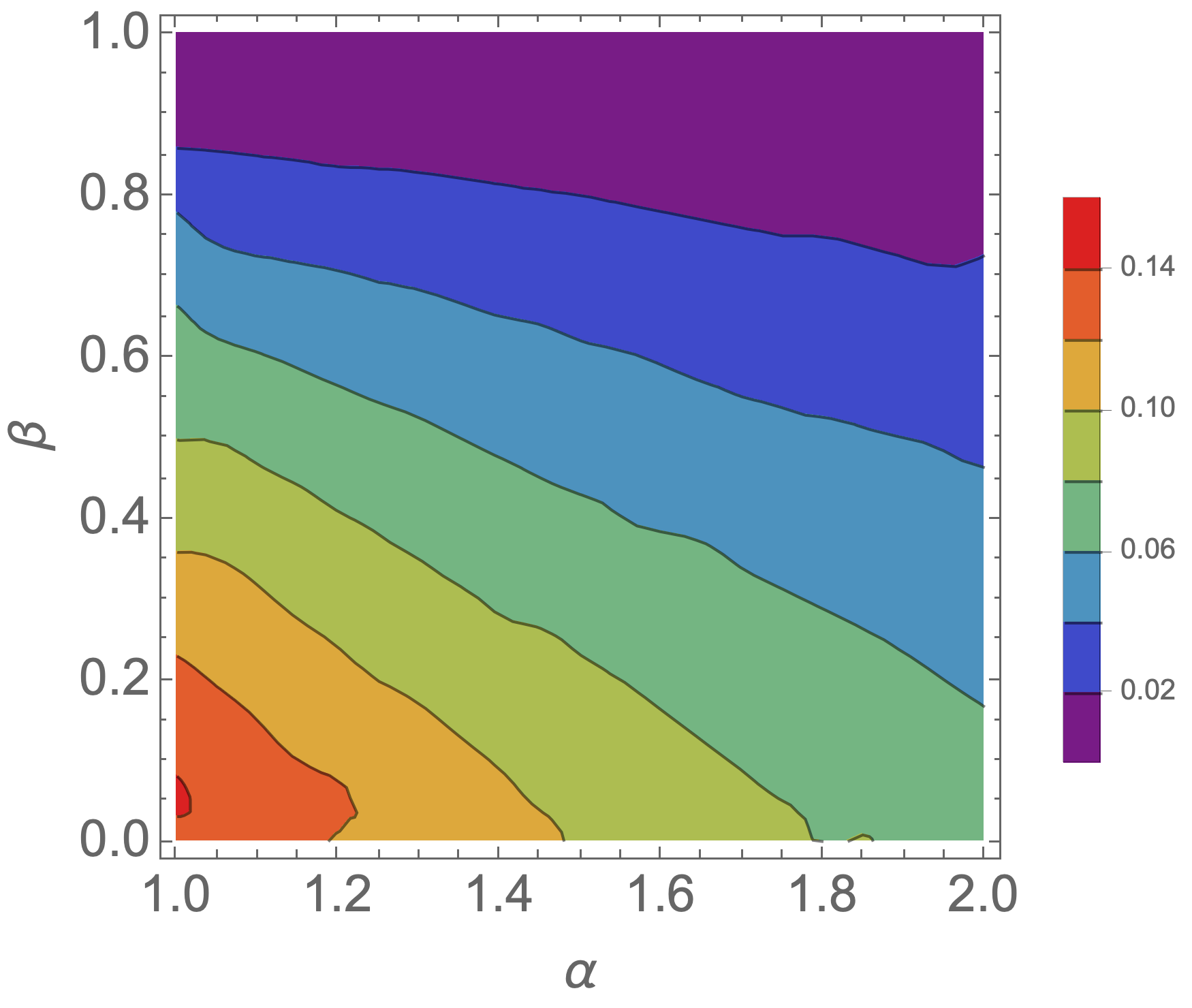}
\caption{{(Top) Computed} $(\alpha,\beta,\epsilon_\text{max})$ points {along with} the resulting {interpolated} surface showing the lifetime $(\epsilon)$ of the limit cycle as a function of $\alpha$ and $\beta$. (Bottom) Contour plot of the {interpolation} function, showing the nonlinear curvature of the bifurcation surface.}
\label{fig:Surface}
\end{figure}

We can numerically approximate the value of $\epsilon$ where the limit cycle disappears for arbitrarily values of $\alpha \in (1,2]$ and $\beta \in (0,{1}]$. To do this, {we} use the following steps:
\begin{enumerate}
\item \textbf{Input:} $\alpha$, $\beta$.
\item \textbf{Initialize:} $\epsilon = \tfrac{1}{100}$. Compute $a = a^* + \epsilon$, where $a^*$ is given by \cref{eqn:astar}.
\item Numerically integrate the time-inverted dynamics:
\begin{equation*}
\dot{u}_i = -u_i\left[f(\mathbf{u} - \bar{f}(\mathbf{u})\right],
\end{equation*}
with $\mathbf{u}_0 = \langle{0.99,0.005,0.005}\rangle$. This accomplishes two things: (i)
The modified dynamics invert the stability of the limit cycle, and (ii) the initial condition starts the trajectory close to the boundary. Let $\mathbf{u}(t)$ be the resulting solution.
 
\item Compute $d(t) = \lVert\mathbf{u}(t) - {{\bar{\mathbf{u}}}}\rVert^2$. 
\item Fit $d(t) \sim \gamma_0 + \gamma_1 t$.
\item If $\gamma_1 < 0$ with $p$-value less than $0.001$, then statistically the trajectory is decaying toward a limit cycle, and we set $\epsilon := \epsilon + \tfrac{1}{1000}$ and go to Step 3. Otherwise, stop and return $\epsilon$.
\end{enumerate}
Using this technique, we can generate an {interpolated} surface showing the dependence of $\epsilon$ on the parameters $\alpha$ and $\beta$ (see \cref{fig:Surface}). Before continuing our analysis, we note that a similar procedure can be used to {find} limit cycles within this dynamical system. Mathematica code to generate all figures is provided in the SI.

{The resolution of the algorithm prevents a complete characterization of $\epsilon_\text{max}$ for all $(\alpha,\beta)$ input pairs. For $\beta$ sufficiently large, the algorithm returns its smallest value, and we can only deduce that $\epsilon_\text{max} < 0.01$. Using this information, we fit the empirically determined $(\alpha,\beta,\epsilon_\text{max})$ points for which $\epsilon_\text{max} > 0.01$ using a generalized linear model. The resulting fit is given by
\begin{multline*}
\hat{\epsilon}_\text{max}(\alpha, \beta) \approx \\  \max\{0.01, 0.224  -0.078 \alpha -0.217 \beta + 0.074 \alpha  \beta\}.
\end{multline*}
}
The {adjusted-$r^2$} of this fit is {$0.99$} showing good explanatory power. The parameter table for the model is 
{
\begin{equation*}
\begin{array}{l|llll}
\text{} & \text{Est.} & \text{SE} & \text{t-Stat} & \text{$p$-val.} \\
\hline
 1 & 0.224 & 0.00155 & 144. & 1.44\times 10^{-301} \\
 \alpha  & -0.0781 & 0.00101 & -77.7 & 4.99\times
   10^{-215} \\
 \beta  & -0.217 & 0.00303 & -71.6 & 5.67\times 10^{-204}
   \\
 \alpha  \beta  & 0.0743 & 0.00199 & 37.4 & 4.59\times
   10^{-121}. \\
\end{array}
\end{equation*}
}
This statistical analysis shows that the maximum period of a limit cycle (correlated to $\epsilon_\text{max}$) is negatively proportional to both $\alpha$ and $\beta$. However, since the interaction term is non-trivial, the interaction between $\alpha$ and $\beta$ can increase the $\epsilon_\text{max}$ before the limit cycle disappears.  {This is consistent with the interpolated surface shown in \cref{fig:Surface}.}

{
\cref{fig:Surface} (top) illustrates {an approximate three-dimensional} bifurcation diagram (for $b = 0$) for the triple $(\alpha,\beta,\epsilon)$. If $\epsilon < 0$ (i.e., $(\alpha,\beta,\epsilon)$ lies below the $\alpha-\beta$ plane), then the interior fixed point is globally unstable. On the other hand, the inputs $(\alpha,\beta,\epsilon)$ generate a limit cycle if $(\alpha,\beta,\epsilon)$ falls between the surface and the $\alpha-\beta$ plane. If  $(\alpha,\beta,\epsilon)$ lies above the surface, then the interior fixed point is globally attracting. This surface is accurate only to the resolution of the algorithm, as we have already noted. The contour plot in \cref{fig:Surface} (bottom) better illustrates the curvature of the bifurcation surface.}

\section{Conclusion}
In this letter, we used the replicator equation to model higher-order interactions between three (or more) species through the introduction of an interaction tensor. We showed that the dynamics that results in this case are fundamentally different from the ordinary binary interactions modelled by the {standard} replicator dynamics with a payoff matrix. In particular, we studied a generalized rock-paper-scissors model and showed the existence of a non-degenerate {subcritical} Hopf bifurcation that allows an {unstable} limit cycle to emerge when higher order interactions are allowed. This is in contrast to classical results on three strategy games in the {standard} replicator equation, in which limit cycles cannot emerge as a result of the degeneracy of a similar Hopf bifurcation \cite{Z80}. 

While this letter provides a framework for modelling higher-order interactions within a replicator framework, there are several possible future directions. Generalizing the interaction rules used to construct the three-way interaction tensor could lead to a generalization of the Folk theorem in certain cases. It also would be interesting to introduce a spatial component as in \cite{CV97,V89,V91,dB13,GMd21} and determine what spatial dynamics emerge as a result of these higher-order interactions. 

\section{Acknowledgements}
C.G. was supported in part by the National Science Foundation under grant DMS-1814876. R.W. was supported by the Talents Grants at Beijing Forestry University.

\bibliography{HORPS}

\begin{thebibliography}{62}%
\makeatletter
\providecommand \@ifxundefined [1]{%
 \@ifx{#1\undefined}
}%
\providecommand \@ifnum [1]{%
 \ifnum #1\expandafter \@firstoftwo
 \else \expandafter \@secondoftwo
 \fi
}%
\providecommand \@ifx [1]{%
 \ifx #1\expandafter \@firstoftwo
 \else \expandafter \@secondoftwo
 \fi
}%
\providecommand \natexlab [1]{#1}%
\providecommand \enquote  [1]{``#1''}%
\providecommand \bibnamefont  [1]{#1}%
\providecommand \bibfnamefont [1]{#1}%
\providecommand \citenamefont [1]{#1}%
\providecommand \href@noop [0]{\@secondoftwo}%
\providecommand \href [0]{\begingroup \@sanitize@url \@href}%
\providecommand \@href[1]{\@@startlink{#1}\@@href}%
\providecommand \@@href[1]{\endgroup#1\@@endlink}%
\providecommand \@sanitize@url [0]{\catcode `\\12\catcode `\$12\catcode
  `\&12\catcode `\#12\catcode `\^12\catcode `\_12\catcode `\%12\relax}%
\providecommand \@@startlink[1]{}%
\providecommand \@@endlink[0]{}%
\providecommand \url  [0]{\begingroup\@sanitize@url \@url }%
\providecommand \@url [1]{\endgroup\@href {#1}{\urlprefix }}%
\providecommand \urlprefix  [0]{URL }%
\providecommand \Eprint [0]{\href }%
\providecommand \doibase [0]{https://doi.org/}%
\providecommand \selectlanguage [0]{\@gobble}%
\providecommand \bibinfo  [0]{\@secondoftwo}%
\providecommand \bibfield  [0]{\@secondoftwo}%
\providecommand \translation [1]{[#1]}%
\providecommand \BibitemOpen [0]{}%
\providecommand \bibitemStop [0]{}%
\providecommand \bibitemNoStop [0]{.\EOS\space}%
\providecommand \EOS [0]{\spacefactor3000\relax}%
\providecommand \BibitemShut  [1]{\csname bibitem#1\endcsname}%
\let\auto@bib@innerbib\@empty
\bibitem [{\citenamefont {May}(1972)}]{M72}%
  \BibitemOpen
  \bibfield  {author} {\bibinfo {author} {\bibfnamefont {R.~M.}\ \bibnamefont
  {May}},\ }\bibfield  {title} {\bibinfo {title} {Will a large complex system
  be stable?},\ }\href@noop {} {\bibfield  {journal} {\bibinfo  {journal}
  {Nature}\ }\textbf {\bibinfo {volume} {238}},\ \bibinfo {pages} {413}
  (\bibinfo {year} {1972})}\BibitemShut {NoStop}%
\bibitem [{\citenamefont {Pomerantz}(1981)}]{P81}%
  \BibitemOpen
  \bibfield  {author} {\bibinfo {author} {\bibfnamefont {M.~J.}\ \bibnamefont
  {Pomerantz}},\ }\bibfield  {title} {\bibinfo {title} {Do" higher order
  interactions" in competition systems really exist?},\ }\href@noop {}
  {\bibfield  {journal} {\bibinfo  {journal} {The American Naturalist}\
  }\textbf {\bibinfo {volume} {117}},\ \bibinfo {pages} {583} (\bibinfo {year}
  {1981})}\BibitemShut {NoStop}%
\bibitem [{\citenamefont {Relyea}\ and\ \citenamefont {Yurewicz}(2002)}]{RY02}%
  \BibitemOpen
  \bibfield  {author} {\bibinfo {author} {\bibfnamefont {R.~A.}\ \bibnamefont
  {Relyea}}\ and\ \bibinfo {author} {\bibfnamefont {K.~L.}\ \bibnamefont
  {Yurewicz}},\ }\bibfield  {title} {\bibinfo {title} {Predicting community
  outcomes from pairwise interactions: integrating density-and trait-mediated
  effects},\ }\href@noop {} {\bibfield  {journal} {\bibinfo  {journal}
  {Oecologia}\ }\textbf {\bibinfo {volume} {131}},\ \bibinfo {pages} {569}
  (\bibinfo {year} {2002})}\BibitemShut {NoStop}%
\bibitem [{\citenamefont {Kodera}\ \emph {et~al.}(2022)\citenamefont {Kodera},
  \citenamefont {Das}, \citenamefont {Gilbert},\ and\ \citenamefont
  {Lutz}}]{KDGL22}%
  \BibitemOpen
  \bibfield  {author} {\bibinfo {author} {\bibfnamefont {S.~M.}\ \bibnamefont
  {Kodera}}, \bibinfo {author} {\bibfnamefont {P.}~\bibnamefont {Das}},
  \bibinfo {author} {\bibfnamefont {J.~A.}\ \bibnamefont {Gilbert}},\ and\
  \bibinfo {author} {\bibfnamefont {H.~L.}\ \bibnamefont {Lutz}},\ }\bibfield
  {title} {\bibinfo {title} {Conceptual strategies for characterizing
  interactions in microbial communities},\ }\href@noop {} {\bibfield  {journal}
  {\bibinfo  {journal} {Iscience}\ ,\ \bibinfo {pages} {103775}} (\bibinfo
  {year} {2022})}\BibitemShut {NoStop}%
\bibitem [{\citenamefont {Weibull}(1997)}]{W97}%
  \BibitemOpen
  \bibfield  {author} {\bibinfo {author} {\bibfnamefont {J.~W.}\ \bibnamefont
  {Weibull}},\ }\href@noop {} {\emph {\bibinfo {title} {Evolutionary Game
  Theory}}}\ (\bibinfo  {publisher} {MIT Press},\ \bibinfo {year}
  {1997})\BibitemShut {NoStop}%
\bibitem [{\citenamefont {Hofbauer}\ and\ \citenamefont
  {Sigmund}(1998)}]{HS98}%
  \BibitemOpen
  \bibfield  {author} {\bibinfo {author} {\bibfnamefont {J.}~\bibnamefont
  {Hofbauer}}\ and\ \bibinfo {author} {\bibfnamefont {K.}~\bibnamefont
  {Sigmund}},\ }\href@noop {} {\emph {\bibinfo {title} {{Evolutionary Games and
  Population Dynamics}}}}\ (\bibinfo  {publisher} {{Cambridge University
  Press}},\ \bibinfo {year} {1998})\BibitemShut {NoStop}%
\bibitem [{\citenamefont {Hofbauer}\ and\ \citenamefont
  {Sigmund}(2003)}]{HS03}%
  \BibitemOpen
  \bibfield  {author} {\bibinfo {author} {\bibfnamefont {J.}~\bibnamefont
  {Hofbauer}}\ and\ \bibinfo {author} {\bibfnamefont {K.}~\bibnamefont
  {Sigmund}},\ }\bibfield  {title} {\bibinfo {title} {{Evolutionary Game
  Dynamics}},\ }\href@noop {} {\bibfield  {journal} {\bibinfo  {journal}
  {Bulletin of the American Mathematical Society}\ }\textbf {\bibinfo {volume}
  {40}},\ \bibinfo {pages} {479} (\bibinfo {year} {2003})}\BibitemShut
  {NoStop}%
\bibitem [{\citenamefont {Levine}\ \emph {et~al.}(2017)\citenamefont {Levine},
  \citenamefont {Bascompte}, \citenamefont {Adler},\ and\ \citenamefont
  {Allesina}}]{LBAA17}%
  \BibitemOpen
  \bibfield  {author} {\bibinfo {author} {\bibfnamefont {J.~M.}\ \bibnamefont
  {Levine}}, \bibinfo {author} {\bibfnamefont {J.}~\bibnamefont {Bascompte}},
  \bibinfo {author} {\bibfnamefont {P.~B.}\ \bibnamefont {Adler}},\ and\
  \bibinfo {author} {\bibfnamefont {S.}~\bibnamefont {Allesina}},\ }\bibfield
  {title} {\bibinfo {title} {Beyond pairwise mechanisms of species coexistence
  in complex communities},\ }\href@noop {} {\bibfield  {journal} {\bibinfo
  {journal} {Nature}\ }\textbf {\bibinfo {volume} {546}},\ \bibinfo {pages}
  {56} (\bibinfo {year} {2017})}\BibitemShut {NoStop}%
\bibitem [{\citenamefont {Grilli}\ \emph {et~al.}(2017)\citenamefont {Grilli},
  \citenamefont {Barab{\'a}s}, \citenamefont {Michalska-Smith},\ and\
  \citenamefont {Allesina}}]{GBMA17}%
  \BibitemOpen
  \bibfield  {author} {\bibinfo {author} {\bibfnamefont {J.}~\bibnamefont
  {Grilli}}, \bibinfo {author} {\bibfnamefont {G.}~\bibnamefont {Barab{\'a}s}},
  \bibinfo {author} {\bibfnamefont {M.~J.}\ \bibnamefont {Michalska-Smith}},\
  and\ \bibinfo {author} {\bibfnamefont {S.}~\bibnamefont {Allesina}},\
  }\bibfield  {title} {\bibinfo {title} {Higher-order interactions stabilize
  dynamics in competitive network models},\ }\href@noop {} {\bibfield
  {journal} {\bibinfo  {journal} {Nature}\ }\textbf {\bibinfo {volume} {548}},\
  \bibinfo {pages} {210} (\bibinfo {year} {2017})}\BibitemShut {NoStop}%
\bibitem [{\citenamefont {Bairey}\ \emph {et~al.}(2016)\citenamefont {Bairey},
  \citenamefont {Kelsic},\ and\ \citenamefont {Kishony}}]{BKK16}%
  \BibitemOpen
  \bibfield  {author} {\bibinfo {author} {\bibfnamefont {E.}~\bibnamefont
  {Bairey}}, \bibinfo {author} {\bibfnamefont {E.~D.}\ \bibnamefont {Kelsic}},\
  and\ \bibinfo {author} {\bibfnamefont {R.}~\bibnamefont {Kishony}},\
  }\bibfield  {title} {\bibinfo {title} {High-order species interactions shape
  ecosystem diversity},\ }\href@noop {} {\bibfield  {journal} {\bibinfo
  {journal} {Nature communications}\ }\textbf {\bibinfo {volume} {7}},\
  \bibinfo {pages} {1} (\bibinfo {year} {2016})}\BibitemShut {NoStop}%
\bibitem [{\citenamefont {McClean}\ \emph {et~al.}(2019)\citenamefont
  {McClean}, \citenamefont {Friman}, \citenamefont {Finn}, \citenamefont
  {Salzberg},\ and\ \citenamefont {Donohue}}]{MFFS19}%
  \BibitemOpen
  \bibfield  {author} {\bibinfo {author} {\bibfnamefont {D.}~\bibnamefont
  {McClean}}, \bibinfo {author} {\bibfnamefont {V.-P.}\ \bibnamefont {Friman}},
  \bibinfo {author} {\bibfnamefont {A.}~\bibnamefont {Finn}}, \bibinfo {author}
  {\bibfnamefont {L.~I.}\ \bibnamefont {Salzberg}},\ and\ \bibinfo {author}
  {\bibfnamefont {I.}~\bibnamefont {Donohue}},\ }\bibfield  {title} {\bibinfo
  {title} {Coping with multiple enemies: pairwise interactions do not predict
  evolutionary change in complex multitrophic communities},\ }\href@noop {}
  {\bibfield  {journal} {\bibinfo  {journal} {Oikos}\ }\textbf {\bibinfo
  {volume} {128}},\ \bibinfo {pages} {1588} (\bibinfo {year}
  {2019})}\BibitemShut {NoStop}%
\bibitem [{\citenamefont {Skardal}\ \emph {et~al.}(2021)\citenamefont
  {Skardal}, \citenamefont {Arola-Fern{\'a}ndez}, \citenamefont {Taylor},\ and\
  \citenamefont {Arenas}}]{SATA21}%
  \BibitemOpen
  \bibfield  {author} {\bibinfo {author} {\bibfnamefont {P.~S.}\ \bibnamefont
  {Skardal}}, \bibinfo {author} {\bibfnamefont {L.}~\bibnamefont
  {Arola-Fern{\'a}ndez}}, \bibinfo {author} {\bibfnamefont {D.}~\bibnamefont
  {Taylor}},\ and\ \bibinfo {author} {\bibfnamefont {A.}~\bibnamefont
  {Arenas}},\ }\bibfield  {title} {\bibinfo {title} {Higher-order interactions
  can better optimize network synchronization},\ }\href@noop {} {\bibfield
  {journal} {\bibinfo  {journal} {Physical Review Research}\ }\textbf {\bibinfo
  {volume} {3}},\ \bibinfo {pages} {043193} (\bibinfo {year}
  {2021})}\BibitemShut {NoStop}%
\bibitem [{\citenamefont {Kleinhesselink}\ \emph {et~al.}(2022)\citenamefont
  {Kleinhesselink}, \citenamefont {Kraft}, \citenamefont {Pacala},\ and\
  \citenamefont {Levine}}]{KKPL22}%
  \BibitemOpen
  \bibfield  {author} {\bibinfo {author} {\bibfnamefont {A.~R.}\ \bibnamefont
  {Kleinhesselink}}, \bibinfo {author} {\bibfnamefont {N.~J.}\ \bibnamefont
  {Kraft}}, \bibinfo {author} {\bibfnamefont {S.~W.}\ \bibnamefont {Pacala}},\
  and\ \bibinfo {author} {\bibfnamefont {J.~M.}\ \bibnamefont {Levine}},\
  }\bibfield  {title} {\bibinfo {title} {Detecting and interpreting
  higher-order interactions in ecological communities},\ }\href@noop {}
  {\bibfield  {journal} {\bibinfo  {journal} {Ecology letters}\ }\textbf
  {\bibinfo {volume} {25}},\ \bibinfo {pages} {1604} (\bibinfo {year}
  {2022})}\BibitemShut {NoStop}%
\bibitem [{\citenamefont {Gibbs}\ \emph {et~al.}(2022)\citenamefont {Gibbs},
  \citenamefont {Levin},\ and\ \citenamefont {Levine}}]{GLL22}%
  \BibitemOpen
  \bibfield  {author} {\bibinfo {author} {\bibfnamefont {T.}~\bibnamefont
  {Gibbs}}, \bibinfo {author} {\bibfnamefont {S.~A.}\ \bibnamefont {Levin}},\
  and\ \bibinfo {author} {\bibfnamefont {J.~M.}\ \bibnamefont {Levine}},\
  }\bibfield  {title} {\bibinfo {title} {Coexistence in diverse communities
  with higher-order interactions},\ }\href
  {https://doi.org/10.1073/pnas.2205063119} {\bibfield  {journal} {\bibinfo
  {journal} {Proceedings of the National Academy of Sciences}\ }\textbf
  {\bibinfo {volume} {119}},\ \bibinfo {pages} {e2205063119} (\bibinfo {year}
  {2022})},\ \Eprint
  {https://arxiv.org/abs/https://www.pnas.org/doi/pdf/10.1073/pnas.2205063119}
  {https://www.pnas.org/doi/pdf/10.1073/pnas.2205063119} \BibitemShut {NoStop}%
\bibitem [{\citenamefont {Battiston}\ \emph {et~al.}(2021)\citenamefont
  {Battiston}, \citenamefont {Amico}, \citenamefont {Barrat}, \citenamefont
  {Bianconi}, \citenamefont {Ferraz~de Arruda}, \citenamefont {Franceschiello},
  \citenamefont {Iacopini}, \citenamefont {K{\'e}fi}, \citenamefont {Latora},
  \citenamefont {Moreno} \emph {et~al.}}]{BABB21}%
  \BibitemOpen
  \bibfield  {author} {\bibinfo {author} {\bibfnamefont {F.}~\bibnamefont
  {Battiston}}, \bibinfo {author} {\bibfnamefont {E.}~\bibnamefont {Amico}},
  \bibinfo {author} {\bibfnamefont {A.}~\bibnamefont {Barrat}}, \bibinfo
  {author} {\bibfnamefont {G.}~\bibnamefont {Bianconi}}, \bibinfo {author}
  {\bibfnamefont {G.}~\bibnamefont {Ferraz~de Arruda}}, \bibinfo {author}
  {\bibfnamefont {B.}~\bibnamefont {Franceschiello}}, \bibinfo {author}
  {\bibfnamefont {I.}~\bibnamefont {Iacopini}}, \bibinfo {author}
  {\bibfnamefont {S.}~\bibnamefont {K{\'e}fi}}, \bibinfo {author}
  {\bibfnamefont {V.}~\bibnamefont {Latora}}, \bibinfo {author} {\bibfnamefont
  {Y.}~\bibnamefont {Moreno}}, \emph {et~al.},\ }\bibfield  {title} {\bibinfo
  {title} {The physics of higher-order interactions in complex systems},\
  }\href@noop {} {\bibfield  {journal} {\bibinfo  {journal} {Nature Physics}\
  }\textbf {\bibinfo {volume} {17}},\ \bibinfo {pages} {1093} (\bibinfo {year}
  {2021})}\BibitemShut {NoStop}%
\bibitem [{\citenamefont {Battiston}\ \emph {et~al.}(2020)\citenamefont
  {Battiston}, \citenamefont {Cencetti}, \citenamefont {Iacopini},
  \citenamefont {Latora}, \citenamefont {Lucas}, \citenamefont {Patania},
  \citenamefont {Young},\ and\ \citenamefont {Petri}}]{BCIL20}%
  \BibitemOpen
  \bibfield  {author} {\bibinfo {author} {\bibfnamefont {F.}~\bibnamefont
  {Battiston}}, \bibinfo {author} {\bibfnamefont {G.}~\bibnamefont {Cencetti}},
  \bibinfo {author} {\bibfnamefont {I.}~\bibnamefont {Iacopini}}, \bibinfo
  {author} {\bibfnamefont {V.}~\bibnamefont {Latora}}, \bibinfo {author}
  {\bibfnamefont {M.}~\bibnamefont {Lucas}}, \bibinfo {author} {\bibfnamefont
  {A.}~\bibnamefont {Patania}}, \bibinfo {author} {\bibfnamefont {J.-G.}\
  \bibnamefont {Young}},\ and\ \bibinfo {author} {\bibfnamefont
  {G.}~\bibnamefont {Petri}},\ }\bibfield  {title} {\bibinfo {title} {Networks
  beyond pairwise interactions: structure and dynamics},\ }\href@noop {}
  {\bibfield  {journal} {\bibinfo  {journal} {Physics Reports}\ }\textbf
  {\bibinfo {volume} {874}},\ \bibinfo {pages} {1} (\bibinfo {year}
  {2020})}\BibitemShut {NoStop}%
\bibitem [{\citenamefont {Lambiotte}\ \emph {et~al.}(2019)\citenamefont
  {Lambiotte}, \citenamefont {Rosvall},\ and\ \citenamefont
  {Scholtes}}]{LRS19}%
  \BibitemOpen
  \bibfield  {author} {\bibinfo {author} {\bibfnamefont {R.}~\bibnamefont
  {Lambiotte}}, \bibinfo {author} {\bibfnamefont {M.}~\bibnamefont {Rosvall}},\
  and\ \bibinfo {author} {\bibfnamefont {I.}~\bibnamefont {Scholtes}},\
  }\bibfield  {title} {\bibinfo {title} {From networks to optimal higher-order
  models of complex systems},\ }\href@noop {} {\bibfield  {journal} {\bibinfo
  {journal} {Nature physics}\ }\textbf {\bibinfo {volume} {15}},\ \bibinfo
  {pages} {313} (\bibinfo {year} {2019})}\BibitemShut {NoStop}%
\bibitem [{\citenamefont {Swain}\ \emph {et~al.}(2022)\citenamefont {Swain},
  \citenamefont {Fussell},\ and\ \citenamefont {Fagan}}]{SFF22}%
  \BibitemOpen
  \bibfield  {author} {\bibinfo {author} {\bibfnamefont {A.}~\bibnamefont
  {Swain}}, \bibinfo {author} {\bibfnamefont {L.}~\bibnamefont {Fussell}},\
  and\ \bibinfo {author} {\bibfnamefont {W.~F.}\ \bibnamefont {Fagan}},\
  }\bibfield  {title} {\bibinfo {title} {Higher-order effects, continuous
  species interactions, and trait evolution shape microbial spatial dynamics},\
  }\href@noop {} {\bibfield  {journal} {\bibinfo  {journal} {Proceedings of the
  National Academy of Sciences}\ }\textbf {\bibinfo {volume} {119}},\ \bibinfo
  {pages} {e2020956119} (\bibinfo {year} {2022})}\BibitemShut {NoStop}%
\bibitem [{\citenamefont {Mayfield}\ and\ \citenamefont
  {Stouffer}(2017)}]{MS17}%
  \BibitemOpen
  \bibfield  {author} {\bibinfo {author} {\bibfnamefont {M.~M.}\ \bibnamefont
  {Mayfield}}\ and\ \bibinfo {author} {\bibfnamefont {D.~B.}\ \bibnamefont
  {Stouffer}},\ }\bibfield  {title} {\bibinfo {title} {Higher-order
  interactions capture unexplained complexity in diverse communities},\
  }\href@noop {} {\bibfield  {journal} {\bibinfo  {journal} {Nature ecology \&
  evolution}\ }\textbf {\bibinfo {volume} {1}},\ \bibinfo {pages} {1} (\bibinfo
  {year} {2017})}\BibitemShut {NoStop}%
\bibitem [{\citenamefont {Mickalide}\ and\ \citenamefont {Kuehn}(2019)}]{MK19}%
  \BibitemOpen
  \bibfield  {author} {\bibinfo {author} {\bibfnamefont {H.}~\bibnamefont
  {Mickalide}}\ and\ \bibinfo {author} {\bibfnamefont {S.}~\bibnamefont
  {Kuehn}},\ }\bibfield  {title} {\bibinfo {title} {Higher-order interaction
  between species inhibits bacterial invasion of a phototroph-predator
  microbial community},\ }\href@noop {} {\bibfield  {journal} {\bibinfo
  {journal} {Cell Systems}\ }\textbf {\bibinfo {volume} {9}},\ \bibinfo {pages}
  {521} (\bibinfo {year} {2019})}\BibitemShut {NoStop}%
\bibitem [{\citenamefont {Deng}\ \emph {et~al.}(2022)\citenamefont {Deng},
  \citenamefont {Taylor},\ and\ \citenamefont {Saavedra}}]{DTS22}%
  \BibitemOpen
  \bibfield  {author} {\bibinfo {author} {\bibfnamefont {J.}~\bibnamefont
  {Deng}}, \bibinfo {author} {\bibfnamefont {W.}~\bibnamefont {Taylor}},\ and\
  \bibinfo {author} {\bibfnamefont {S.}~\bibnamefont {Saavedra}},\ }\bibfield
  {title} {\bibinfo {title} {Understanding the impact of third-party species on
  pairwise coexistence},\ }\href@noop {} {\bibfield  {journal} {\bibinfo
  {journal} {PLOS Computational Biology}\ }\textbf {\bibinfo {volume} {18}},\
  \bibinfo {pages} {e1010630} (\bibinfo {year} {2022})}\BibitemShut {NoStop}%
\bibitem [{\citenamefont {Hofbauer}\ \emph {et~al.}(1982)\citenamefont
  {Hofbauer}, \citenamefont {Schuster},\ and\ \citenamefont {Sigmund}}]{HSS82}%
  \BibitemOpen
  \bibfield  {author} {\bibinfo {author} {\bibfnamefont {J.}~\bibnamefont
  {Hofbauer}}, \bibinfo {author} {\bibfnamefont {P.}~\bibnamefont {Schuster}},\
  and\ \bibinfo {author} {\bibfnamefont {K.}~\bibnamefont {Sigmund}},\
  }\bibfield  {title} {\bibinfo {title} {Game dynamics in mendelian
  populations},\ }\href@noop {} {\bibfield  {journal} {\bibinfo  {journal}
  {Biological Cybernetics}\ }\textbf {\bibinfo {volume} {43}},\ \bibinfo
  {pages} {51} (\bibinfo {year} {1982})}\BibitemShut {NoStop}%
\bibitem [{\citenamefont {Gokhale}\ and\ \citenamefont
  {Traulsen}(2010)}]{GT10}%
  \BibitemOpen
  \bibfield  {author} {\bibinfo {author} {\bibfnamefont {C.~S.}\ \bibnamefont
  {Gokhale}}\ and\ \bibinfo {author} {\bibfnamefont {A.}~\bibnamefont
  {Traulsen}},\ }\bibfield  {title} {\bibinfo {title} {Evolutionary games in
  the multiverse},\ }\href@noop {} {\bibfield  {journal} {\bibinfo  {journal}
  {Proceedings of the National Academy of Sciences}\ }\textbf {\bibinfo
  {volume} {107}},\ \bibinfo {pages} {5500} (\bibinfo {year}
  {2010})}\BibitemShut {NoStop}%
\bibitem [{\citenamefont {Zhang}\ \emph {et~al.}(2022)\citenamefont {Zhang},
  \citenamefont {Peng}, \citenamefont {Zhou}, \citenamefont {Wang},\ and\
  \citenamefont {Li}}]{ZPZW22}%
  \BibitemOpen
  \bibfield  {author} {\bibinfo {author} {\bibfnamefont {X.}~\bibnamefont
  {Zhang}}, \bibinfo {author} {\bibfnamefont {P.}~\bibnamefont {Peng}},
  \bibinfo {author} {\bibfnamefont {Y.}~\bibnamefont {Zhou}}, \bibinfo {author}
  {\bibfnamefont {H.}~\bibnamefont {Wang}},\ and\ \bibinfo {author}
  {\bibfnamefont {W.}~\bibnamefont {Li}},\ }\bibfield  {title} {\bibinfo
  {title} {Evolutionary game-theoretical analysis for general multiplayer
  asymmetric games},\ }\href@noop {} {\bibfield  {journal} {\bibinfo  {journal}
  {arXiv preprint arXiv:2206.11114}\ } (\bibinfo {year} {2022})}\BibitemShut
  {NoStop}%
\bibitem [{\citenamefont {Peixe}\ and\ \citenamefont
  {Rodrigues}(2022)}]{PR22a}%
  \BibitemOpen
  \bibfield  {author} {\bibinfo {author} {\bibfnamefont {T.}~\bibnamefont
  {Peixe}}\ and\ \bibinfo {author} {\bibfnamefont {A.}~\bibnamefont
  {Rodrigues}},\ }\bibfield  {title} {\bibinfo {title} {Persistent strange
  attractors in 3d polymatrix replicators},\ }\href@noop {} {\bibfield
  {journal} {\bibinfo  {journal} {Physica D: Nonlinear Phenomena}\ }\textbf
  {\bibinfo {volume} {438}},\ \bibinfo {pages} {133346} (\bibinfo {year}
  {2022})}\BibitemShut {NoStop}%
\bibitem [{\citenamefont {Alishah}\ and\ \citenamefont {Duarte}(2015)}]{AD15}%
  \BibitemOpen
  \bibfield  {author} {\bibinfo {author} {\bibfnamefont {H.~N.}\ \bibnamefont
  {Alishah}}\ and\ \bibinfo {author} {\bibfnamefont {P.}~\bibnamefont
  {Duarte}},\ }\bibfield  {title} {\bibinfo {title} {Hamiltonian evolutionary
  games},\ }\href {https://doi.org/10.3934/jdg.2015.2.33} {\bibfield  {journal}
  {\bibinfo  {journal} {Journal of Dynamics and Games}\ }\textbf {\bibinfo
  {volume} {2}},\ \bibinfo {pages} {33} (\bibinfo {year} {2015})}\BibitemShut
  {NoStop}%
\bibitem [{\citenamefont {Paulson}\ and\ \citenamefont {Griffin}(2016)}]{PG16}%
  \BibitemOpen
  \bibfield  {author} {\bibinfo {author} {\bibfnamefont {E.}~\bibnamefont
  {Paulson}}\ and\ \bibinfo {author} {\bibfnamefont {C.}~\bibnamefont
  {Griffin}},\ }\bibfield  {title} {\bibinfo {title} {Cooperation can emerge in
  prisoner's dilemma from a multi-species predator prey replicator dynamic},\
  }\href@noop {} {\bibfield  {journal} {\bibinfo  {journal} {Mathematical
  biosciences}\ }\textbf {\bibinfo {volume} {278}},\ \bibinfo {pages} {56}
  (\bibinfo {year} {2016})}\BibitemShut {NoStop}%
\bibitem [{\citenamefont {May}\ and\ \citenamefont {Leonard}(1975)}]{ML75}%
  \BibitemOpen
  \bibfield  {author} {\bibinfo {author} {\bibfnamefont {R.~M.}\ \bibnamefont
  {May}}\ and\ \bibinfo {author} {\bibfnamefont {W.~J.}\ \bibnamefont
  {Leonard}},\ }\bibfield  {title} {\bibinfo {title} {Nonlinear aspects of
  competition between three species},\ }\href@noop {} {\bibfield  {journal}
  {\bibinfo  {journal} {SIAM journal on applied mathematics}\ }\textbf
  {\bibinfo {volume} {29}},\ \bibinfo {pages} {243} (\bibinfo {year}
  {1975})}\BibitemShut {NoStop}%
\bibitem [{\citenamefont {Mobilia}(2010)}]{M10}%
  \BibitemOpen
  \bibfield  {author} {\bibinfo {author} {\bibfnamefont {M.}~\bibnamefont
  {Mobilia}},\ }\bibfield  {title} {\bibinfo {title} {Oscillatory dynamics in
  rock--paper--scissors games with mutations},\ }\href@noop {} {\bibfield
  {journal} {\bibinfo  {journal} {Journal of Theoretical Biology}\ }\textbf
  {\bibinfo {volume} {264}},\ \bibinfo {pages} {1} (\bibinfo {year}
  {2010})}\BibitemShut {NoStop}%
\bibitem [{\citenamefont {Postlethwaite}\ and\ \citenamefont
  {Rucklidge}(2019)}]{PR19}%
  \BibitemOpen
  \bibfield  {author} {\bibinfo {author} {\bibfnamefont {C.~M.}\ \bibnamefont
  {Postlethwaite}}\ and\ \bibinfo {author} {\bibfnamefont {A.~M.}\ \bibnamefont
  {Rucklidge}},\ }\bibfield  {title} {\bibinfo {title} {A trio of heteroclinic
  bifurcations arising from a model of spatially-extended
  rock--paper--scissors},\ }\href@noop {} {\bibfield  {journal} {\bibinfo
  {journal} {Nonlinearity}\ }\textbf {\bibinfo {volume} {32}},\ \bibinfo
  {pages} {1375} (\bibinfo {year} {2019})}\BibitemShut {NoStop}%
\bibitem [{\citenamefont {Hua}\ \emph {et~al.}(2013)\citenamefont {Hua},
  \citenamefont {Dai},\ and\ \citenamefont {Lin}}]{HDL13}%
  \BibitemOpen
  \bibfield  {author} {\bibinfo {author} {\bibfnamefont {D.-y.}\ \bibnamefont
  {Hua}}, \bibinfo {author} {\bibfnamefont {L.-c.}\ \bibnamefont {Dai}},\ and\
  \bibinfo {author} {\bibfnamefont {C.}~\bibnamefont {Lin}},\ }\bibfield
  {title} {\bibinfo {title} {Four-and three-state rock-paper-scissors games
  with long-range selection},\ }\href@noop {} {\bibfield  {journal} {\bibinfo
  {journal} {EPL (Europhysics Letters)}\ }\textbf {\bibinfo {volume} {101}},\
  \bibinfo {pages} {38004} (\bibinfo {year} {2013})}\BibitemShut {NoStop}%
\bibitem [{\citenamefont {Szczesny}\ \emph {et~al.}(2014)\citenamefont
  {Szczesny}, \citenamefont {Mobilia},\ and\ \citenamefont
  {Rucklidge}}]{SMR14}%
  \BibitemOpen
  \bibfield  {author} {\bibinfo {author} {\bibfnamefont {B.}~\bibnamefont
  {Szczesny}}, \bibinfo {author} {\bibfnamefont {M.}~\bibnamefont {Mobilia}},\
  and\ \bibinfo {author} {\bibfnamefont {A.~M.}\ \bibnamefont {Rucklidge}},\
  }\bibfield  {title} {\bibinfo {title} {Characterization of spiraling patterns
  in spatial rock-paper-scissors games},\ }\href@noop {} {\bibfield  {journal}
  {\bibinfo  {journal} {Physical Review E}\ }\textbf {\bibinfo {volume} {90}},\
  \bibinfo {pages} {032704} (\bibinfo {year} {2014})}\BibitemShut {NoStop}%
\bibitem [{\citenamefont {Szczesny}\ \emph {et~al.}(2013)\citenamefont
  {Szczesny}, \citenamefont {Mobilia},\ and\ \citenamefont
  {Rucklidge}}]{SMR13}%
  \BibitemOpen
  \bibfield  {author} {\bibinfo {author} {\bibfnamefont {B.}~\bibnamefont
  {Szczesny}}, \bibinfo {author} {\bibfnamefont {M.}~\bibnamefont {Mobilia}},\
  and\ \bibinfo {author} {\bibfnamefont {A.~M.}\ \bibnamefont {Rucklidge}},\
  }\bibfield  {title} {\bibinfo {title} {When does cyclic dominance lead to
  stable spiral waves?},\ }\href@noop {} {\bibfield  {journal} {\bibinfo
  {journal} {EPL (Europhysics Letters)}\ }\textbf {\bibinfo {volume} {102}},\
  \bibinfo {pages} {28012} (\bibinfo {year} {2013})}\BibitemShut {NoStop}%
\bibitem [{\citenamefont {Szolnoki}\ \emph {et~al.}(2014)\citenamefont
  {Szolnoki}, \citenamefont {Mobilia}, \citenamefont {Jiang}, \citenamefont
  {Szczesny}, \citenamefont {Rucklidge},\ and\ \citenamefont {Perc}}]{SMJS14}%
  \BibitemOpen
  \bibfield  {author} {\bibinfo {author} {\bibfnamefont {A.}~\bibnamefont
  {Szolnoki}}, \bibinfo {author} {\bibfnamefont {M.}~\bibnamefont {Mobilia}},
  \bibinfo {author} {\bibfnamefont {L.-L.}\ \bibnamefont {Jiang}}, \bibinfo
  {author} {\bibfnamefont {B.}~\bibnamefont {Szczesny}}, \bibinfo {author}
  {\bibfnamefont {A.~M.}\ \bibnamefont {Rucklidge}},\ and\ \bibinfo {author}
  {\bibfnamefont {M.}~\bibnamefont {Perc}},\ }\bibfield  {title} {\bibinfo
  {title} {Cyclic dominance in evolutionary games: a review},\ }\href@noop {}
  {\bibfield  {journal} {\bibinfo  {journal} {Journal of the Royal Society
  Interface}\ }\textbf {\bibinfo {volume} {11}},\ \bibinfo {pages} {20140735}
  (\bibinfo {year} {2014})}\BibitemShut {NoStop}%
\bibitem [{\citenamefont {Reichenbach}\ \emph {et~al.}(2008)\citenamefont
  {Reichenbach}, \citenamefont {Mobilia},\ and\ \citenamefont {Frey}}]{RMF08}%
  \BibitemOpen
  \bibfield  {author} {\bibinfo {author} {\bibfnamefont {T.}~\bibnamefont
  {Reichenbach}}, \bibinfo {author} {\bibfnamefont {M.}~\bibnamefont
  {Mobilia}},\ and\ \bibinfo {author} {\bibfnamefont {E.}~\bibnamefont
  {Frey}},\ }\bibfield  {title} {\bibinfo {title} {Self-organization of mobile
  populations in cyclic competition},\ }\href@noop {} {\bibfield  {journal}
  {\bibinfo  {journal} {Journal of Theoretical Biology}\ }\textbf {\bibinfo
  {volume} {254}},\ \bibinfo {pages} {368} (\bibinfo {year}
  {2008})}\BibitemShut {NoStop}%
\bibitem [{\citenamefont {Reichenbach}\ \emph {et~al.}(2007)\citenamefont
  {Reichenbach}, \citenamefont {Mobilia},\ and\ \citenamefont {Frey}}]{RMF07}%
  \BibitemOpen
  \bibfield  {author} {\bibinfo {author} {\bibfnamefont {T.}~\bibnamefont
  {Reichenbach}}, \bibinfo {author} {\bibfnamefont {M.}~\bibnamefont
  {Mobilia}},\ and\ \bibinfo {author} {\bibfnamefont {E.}~\bibnamefont
  {Frey}},\ }\bibfield  {title} {\bibinfo {title} {Mobility promotes and
  jeopardizes biodiversity in rock--paper--scissors games},\ }\href@noop {}
  {\bibfield  {journal} {\bibinfo  {journal} {Nature}\ }\textbf {\bibinfo
  {volume} {448}},\ \bibinfo {pages} {1046} (\bibinfo {year}
  {2007})}\BibitemShut {NoStop}%
\bibitem [{\citenamefont {Postlethwaite}\ and\ \citenamefont
  {Rucklidge}(2017)}]{PR17}%
  \BibitemOpen
  \bibfield  {author} {\bibinfo {author} {\bibfnamefont {C.}~\bibnamefont
  {Postlethwaite}}\ and\ \bibinfo {author} {\bibfnamefont {A.}~\bibnamefont
  {Rucklidge}},\ }\bibfield  {title} {\bibinfo {title} {Spirals and
  heteroclinic cycles in a spatially extended rock-paper-scissors model of
  cyclic dominance},\ }\href@noop {} {\bibfield  {journal} {\bibinfo  {journal}
  {EPL (Europhysics Letters)}\ }\textbf {\bibinfo {volume} {117}},\ \bibinfo
  {pages} {48006} (\bibinfo {year} {2017})}\BibitemShut {NoStop}%
\bibitem [{\citenamefont {Bazeia}\ \emph {et~al.}(2017)\citenamefont {Bazeia},
  \citenamefont {Menezes}, \citenamefont {De~Oliveira},\ and\ \citenamefont
  {Ramos}}]{BMDR17}%
  \BibitemOpen
  \bibfield  {author} {\bibinfo {author} {\bibfnamefont {D.}~\bibnamefont
  {Bazeia}}, \bibinfo {author} {\bibfnamefont {J.}~\bibnamefont {Menezes}},
  \bibinfo {author} {\bibfnamefont {B.}~\bibnamefont {De~Oliveira}},\ and\
  \bibinfo {author} {\bibfnamefont {J.}~\bibnamefont {Ramos}},\ }\bibfield
  {title} {\bibinfo {title} {Hamming distance and mobility behavior in
  generalized rock-paper-scissors models},\ }\href@noop {} {\bibfield
  {journal} {\bibinfo  {journal} {EPL (Europhysics Letters)}\ }\textbf
  {\bibinfo {volume} {119}},\ \bibinfo {pages} {58003} (\bibinfo {year}
  {2017})}\BibitemShut {NoStop}%
\bibitem [{\citenamefont {He}\ \emph {et~al.}(2010)\citenamefont {He},
  \citenamefont {Mobilia},\ and\ \citenamefont {T{\"a}uber}}]{HMT10}%
  \BibitemOpen
  \bibfield  {author} {\bibinfo {author} {\bibfnamefont {Q.}~\bibnamefont
  {He}}, \bibinfo {author} {\bibfnamefont {M.}~\bibnamefont {Mobilia}},\ and\
  \bibinfo {author} {\bibfnamefont {U.~C.}\ \bibnamefont {T{\"a}uber}},\
  }\bibfield  {title} {\bibinfo {title} {Spatial rock-paper-scissors models
  with inhomogeneous reaction rates},\ }\href@noop {} {\bibfield  {journal}
  {\bibinfo  {journal} {Physical Review E}\ }\textbf {\bibinfo {volume} {82}},\
  \bibinfo {pages} {051909} (\bibinfo {year} {2010})}\BibitemShut {NoStop}%
\bibitem [{\citenamefont {Kabir}\ and\ \citenamefont {Tanimoto}(2021)}]{KT21}%
  \BibitemOpen
  \bibfield  {author} {\bibinfo {author} {\bibfnamefont {K.~A.}\ \bibnamefont
  {Kabir}}\ and\ \bibinfo {author} {\bibfnamefont {J.}~\bibnamefont
  {Tanimoto}},\ }\bibfield  {title} {\bibinfo {title} {The role of pairwise
  nonlinear evolutionary dynamics in the rock--paper--scissors game with
  noise},\ }\href@noop {} {\bibfield  {journal} {\bibinfo  {journal} {Applied
  Mathematics and Computation}\ }\textbf {\bibinfo {volume} {394}},\ \bibinfo
  {pages} {125767} (\bibinfo {year} {2021})}\BibitemShut {NoStop}%
\bibitem [{\citenamefont {Griffin}\ \emph {et~al.}(2022)\citenamefont
  {Griffin}, \citenamefont {Semonsen},\ and\ \citenamefont {Belmonte}}]{GSB22}%
  \BibitemOpen
  \bibfield  {author} {\bibinfo {author} {\bibfnamefont {C.}~\bibnamefont
  {Griffin}}, \bibinfo {author} {\bibfnamefont {J.}~\bibnamefont {Semonsen}},\
  and\ \bibinfo {author} {\bibfnamefont {A.}~\bibnamefont {Belmonte}},\
  }\bibfield  {title} {\bibinfo {title} {Generalized hamiltonian dynamics and
  chaos in evolutionary games on networks},\ }\href@noop {} {\bibfield
  {journal} {\bibinfo  {journal} {Physica A: Statistical Mechanics and its
  Applications}\ }\textbf {\bibinfo {volume} {597}},\ \bibinfo {pages} {127281}
  (\bibinfo {year} {2022})}\BibitemShut {NoStop}%
\bibitem [{\citenamefont {Postlethwaite}\ and\ \citenamefont
  {Rucklidge}(2022)}]{PR22}%
  \BibitemOpen
  \bibfield  {author} {\bibinfo {author} {\bibfnamefont {C.~M.}\ \bibnamefont
  {Postlethwaite}}\ and\ \bibinfo {author} {\bibfnamefont {A.~M.}\ \bibnamefont
  {Rucklidge}},\ }\bibfield  {title} {\bibinfo {title} {Stability of cycling
  behaviour near a heteroclinic network model of
  rock--paper--scissors--lizard--spock},\ }\href@noop {} {\bibfield  {journal}
  {\bibinfo  {journal} {Nonlinearity}\ }\textbf {\bibinfo {volume} {35}},\
  \bibinfo {pages} {1702} (\bibinfo {year} {2022})}\BibitemShut {NoStop}%
\bibitem [{\citenamefont {Menezes}\ \emph {et~al.}(2019)\citenamefont
  {Menezes}, \citenamefont {Moura},\ and\ \citenamefont {Pereira}}]{MMP19}%
  \BibitemOpen
  \bibfield  {author} {\bibinfo {author} {\bibfnamefont {J.}~\bibnamefont
  {Menezes}}, \bibinfo {author} {\bibfnamefont {B.}~\bibnamefont {Moura}},\
  and\ \bibinfo {author} {\bibfnamefont {T.}~\bibnamefont {Pereira}},\
  }\bibfield  {title} {\bibinfo {title} {Uneven rock-paper-scissors models:
  patterns and coexistence},\ }\href@noop {} {\bibfield  {journal} {\bibinfo
  {journal} {EPL (Europhysics Letters)}\ }\textbf {\bibinfo {volume} {126}},\
  \bibinfo {pages} {18003} (\bibinfo {year} {2019})}\BibitemShut {NoStop}%
\bibitem [{\citenamefont {Avelino}\ \emph {et~al.}(2021)\citenamefont
  {Avelino}, \citenamefont {de~Oliveira},\ and\ \citenamefont
  {Trintin}}]{AOT21}%
  \BibitemOpen
  \bibfield  {author} {\bibinfo {author} {\bibfnamefont {P.}~\bibnamefont
  {Avelino}}, \bibinfo {author} {\bibfnamefont {B.}~\bibnamefont
  {de~Oliveira}},\ and\ \bibinfo {author} {\bibfnamefont {R.}~\bibnamefont
  {Trintin}},\ }\bibfield  {title} {\bibinfo {title} {Weak species in
  rock-paper-scissors models},\ }\href@noop {} {\bibfield  {journal} {\bibinfo
  {journal} {EPL (Europhysics Letters)}\ }\textbf {\bibinfo {volume} {134}},\
  \bibinfo {pages} {48001} (\bibinfo {year} {2021})}\BibitemShut {NoStop}%
\bibitem [{\citenamefont {Park}(2019)}]{P19}%
  \BibitemOpen
  \bibfield  {author} {\bibinfo {author} {\bibfnamefont {J.}~\bibnamefont
  {Park}},\ }\bibfield  {title} {\bibinfo {title} {Fitness-based mutation in
  the spatial rock-paper-scissors game: Shifting of critical mobility for
  extinction},\ }\href@noop {} {\bibfield  {journal} {\bibinfo  {journal} {EPL
  (Europhysics Letters)}\ }\textbf {\bibinfo {volume} {126}},\ \bibinfo {pages}
  {38004} (\bibinfo {year} {2019})}\BibitemShut {NoStop}%
\bibitem [{\citenamefont {Mugnaine}\ \emph {et~al.}(2019)\citenamefont
  {Mugnaine}, \citenamefont {Andrade}, \citenamefont {Szezech},\ and\
  \citenamefont {Bazeia}}]{MASB19}%
  \BibitemOpen
  \bibfield  {author} {\bibinfo {author} {\bibfnamefont {M.}~\bibnamefont
  {Mugnaine}}, \bibinfo {author} {\bibfnamefont {F.}~\bibnamefont {Andrade}},
  \bibinfo {author} {\bibfnamefont {J.}~\bibnamefont {Szezech}},\ and\ \bibinfo
  {author} {\bibfnamefont {D.}~\bibnamefont {Bazeia}},\ }\bibfield  {title}
  {\bibinfo {title} {Basin entropy behavior in a cyclic model of the
  rock-paper-scissors type},\ }\href@noop {} {\bibfield  {journal} {\bibinfo
  {journal} {EPL (Europhysics Letters)}\ }\textbf {\bibinfo {volume} {125}},\
  \bibinfo {pages} {58003} (\bibinfo {year} {2019})}\BibitemShut {NoStop}%
\bibitem [{\citenamefont {Menezes}\ \emph {et~al.}(2022)\citenamefont
  {Menezes}, \citenamefont {Batista}, \citenamefont {Tenorio}, \citenamefont
  {Triaca},\ and\ \citenamefont {Moura}}]{MBTT22}%
  \BibitemOpen
  \bibfield  {author} {\bibinfo {author} {\bibfnamefont {J.}~\bibnamefont
  {Menezes}}, \bibinfo {author} {\bibfnamefont {S.}~\bibnamefont {Batista}},
  \bibinfo {author} {\bibfnamefont {M.}~\bibnamefont {Tenorio}}, \bibinfo
  {author} {\bibfnamefont {E.}~\bibnamefont {Triaca}},\ and\ \bibinfo {author}
  {\bibfnamefont {B.}~\bibnamefont {Moura}},\ }\bibfield  {title} {\bibinfo
  {title} {How local antipredator response unbalances the rock-paper-scissors
  model},\ }\href@noop {} {\bibfield  {journal} {\bibinfo  {journal} {Chaos: An
  Interdisciplinary Journal of Nonlinear Science}\ }\textbf {\bibinfo {volume}
  {32}},\ \bibinfo {pages} {123142} (\bibinfo {year} {2022})}\BibitemShut
  {NoStop}%
\bibitem [{\citenamefont {Itoh}(1987)}]{I87}%
  \BibitemOpen
  \bibfield  {author} {\bibinfo {author} {\bibfnamefont {Y.}~\bibnamefont
  {Itoh}},\ }\bibfield  {title} {\bibinfo {title} {Integrals of a
  lotka-volterra system of odd number of variables},\ }\href@noop {} {\bibfield
   {journal} {\bibinfo  {journal} {Progress of theoretical physics}\ }\textbf
  {\bibinfo {volume} {78}},\ \bibinfo {pages} {507} (\bibinfo {year}
  {1987})}\BibitemShut {NoStop}%
\bibitem [{\citenamefont {Bogoyavlensky}(1988)}]{B88}%
  \BibitemOpen
  \bibfield  {author} {\bibinfo {author} {\bibfnamefont {O.}~\bibnamefont
  {Bogoyavlensky}},\ }\bibfield  {title} {\bibinfo {title} {Integrable
  discretizations of the kdv equation},\ }\href@noop {} {\bibfield  {journal}
  {\bibinfo  {journal} {Physics Letters A}\ }\textbf {\bibinfo {volume}
  {134}},\ \bibinfo {pages} {34} (\bibinfo {year} {1988})}\BibitemShut
  {NoStop}%
\bibitem [{\citenamefont {Veselov}\ and\ \citenamefont {Shabat}(1993)}]{VS93}%
  \BibitemOpen
  \bibfield  {author} {\bibinfo {author} {\bibfnamefont {A.~P.}\ \bibnamefont
  {Veselov}}\ and\ \bibinfo {author} {\bibfnamefont {A.~B.}\ \bibnamefont
  {Shabat}},\ }\bibfield  {title} {\bibinfo {title} {Dressing chains and
  spectral theory of the schrodinger operator},\ }\href@noop {} {\bibfield
  {journal} {\bibinfo  {journal} {Funktsional'nyi Analiz i ego Prilozheniya}\
  }\textbf {\bibinfo {volume} {27}},\ \bibinfo {pages} {1} (\bibinfo {year}
  {1993})}\BibitemShut {NoStop}%
\bibitem [{\citenamefont {Griffin}(2021)}]{G21}%
  \BibitemOpen
  \bibfield  {author} {\bibinfo {author} {\bibfnamefont {C.}~\bibnamefont
  {Griffin}},\ }\bibfield  {title} {\bibinfo {title} {The replicator dynamics
  of zero-sum games arise from a novel poisson algebra},\ }\href@noop {}
  {\bibfield  {journal} {\bibinfo  {journal} {Chaos, Solitons \& Fractals}\
  }\textbf {\bibinfo {volume} {153}},\ \bibinfo {pages} {111508} (\bibinfo
  {year} {2021})}\BibitemShut {NoStop}%
\bibitem [{\citenamefont {Paik}\ and\ \citenamefont {Griffin}(2022)}]{PG22}%
  \BibitemOpen
  \bibfield  {author} {\bibinfo {author} {\bibfnamefont {J.}~\bibnamefont
  {Paik}}\ and\ \bibinfo {author} {\bibfnamefont {C.}~\bibnamefont {Griffin}},\
  }\bibfield  {title} {\bibinfo {title} {Completely integrable replicator
  dynamics associated to competitive networks},\ }\href@noop {} {\bibfield
  {journal} {\bibinfo  {journal} {arXiv preprint arXiv:2211.06501}\ } (\bibinfo
  {year} {2022})}\BibitemShut {NoStop}%
\bibitem [{\citenamefont {Griffin}\ \emph {et~al.}(2021)\citenamefont
  {Griffin}, \citenamefont {Mummah},\ and\ \citenamefont {deForest}}]{GMd21}%
  \BibitemOpen
  \bibfield  {author} {\bibinfo {author} {\bibfnamefont {C.}~\bibnamefont
  {Griffin}}, \bibinfo {author} {\bibfnamefont {R.}~\bibnamefont {Mummah}},\
  and\ \bibinfo {author} {\bibfnamefont {R.}~\bibnamefont {deForest}},\
  }\bibfield  {title} {\bibinfo {title} {A finite population destroys a
  traveling wave in spatial replicator dynamics},\ }\href@noop {} {\bibfield
  {journal} {\bibinfo  {journal} {Chaos, Solitons \& Fractals}\ }\textbf
  {\bibinfo {volume} {146}},\ \bibinfo {pages} {110847} (\bibinfo {year}
  {2021})}\BibitemShut {NoStop}%
\bibitem [{\citenamefont {Zeeman}(1980)}]{Z80}%
  \BibitemOpen
  \bibfield  {author} {\bibinfo {author} {\bibfnamefont {E.~C.}\ \bibnamefont
  {Zeeman}},\ }\bibfield  {title} {\bibinfo {title} {Population dynamics from
  game theory},\ }in\ \href@noop {} {\emph {\bibinfo {booktitle} {Global Theory
  of Dynamical Systems}}},\ \bibinfo {series and number} {\bibinfo {series}
  {Springer Lecture Notes in Mathematics}\ No.\ \bibinfo {number} {819}}\
  (\bibinfo  {publisher} {Springer},\ \bibinfo {year} {1980})\ pp.\ \bibinfo
  {pages} {471--497}\BibitemShut {NoStop}%
\bibitem [{\citenamefont {Feng}\ \emph {et~al.}(2023)\citenamefont {Feng},
  \citenamefont {Liu}, \citenamefont {He}, \citenamefont {Jin}, \citenamefont
  {Wang}, \citenamefont {Yin}, \citenamefont {Cao}, \citenamefont {Huang},
  \citenamefont {Griffin},\ and\ \citenamefont {Wu}}]{FLHJ23}%
  \BibitemOpen
  \bibfield  {author} {\bibinfo {author} {\bibfnamefont {L.}~\bibnamefont
  {Feng}}, \bibinfo {author} {\bibfnamefont {X.}~\bibnamefont {Liu}}, \bibinfo
  {author} {\bibfnamefont {X.}~\bibnamefont {He}}, \bibinfo {author}
  {\bibfnamefont {Y.}~\bibnamefont {Jin}}, \bibinfo {author} {\bibfnamefont
  {J.}~\bibnamefont {Wang}}, \bibinfo {author} {\bibfnamefont {J.}~\bibnamefont
  {Yin}}, \bibinfo {author} {\bibfnamefont {X.}~\bibnamefont {Cao}}, \bibinfo
  {author} {\bibfnamefont {H.}~\bibnamefont {Huang}}, \bibinfo {author}
  {\bibfnamefont {C.}~\bibnamefont {Griffin}},\ and\ \bibinfo {author}
  {\bibfnamefont {R.}~\bibnamefont {Wu}},\ }\bibfield  {title} {\bibinfo
  {title} {Ecological statistical mechanics of high-order interactions in
  complex communities},\ }\href@noop {} {\bibfield  {journal} {\bibinfo
  {journal} {Submitted}\ } (\bibinfo {year} {2023})}\BibitemShut {NoStop}%
\bibitem [{\citenamefont {Griffin}\ \emph {et~al.}(2020)\citenamefont
  {Griffin}, \citenamefont {Jiang},\ and\ \citenamefont {Wu}}]{GJW20}%
  \BibitemOpen
  \bibfield  {author} {\bibinfo {author} {\bibfnamefont {C.}~\bibnamefont
  {Griffin}}, \bibinfo {author} {\bibfnamefont {L.}~\bibnamefont {Jiang}},\
  and\ \bibinfo {author} {\bibfnamefont {R.}~\bibnamefont {Wu}},\ }\bibfield
  {title} {\bibinfo {title} {Analysis of quasi-dynamic ordinary differential
  equations and the quasi-dynamic replicator},\ }\href@noop {} {\bibfield
  {journal} {\bibinfo  {journal} {Physica A: Statistical Mechanics and its
  Applications}\ }\textbf {\bibinfo {volume} {555}},\ \bibinfo {pages} {124422}
  (\bibinfo {year} {2020})}\BibitemShut {NoStop}%
\bibitem [{\citenamefont {Verhulst}(2006)}]{V06}%
  \BibitemOpen
  \bibfield  {author} {\bibinfo {author} {\bibfnamefont {F.}~\bibnamefont
  {Verhulst}},\ }\href@noop {} {\emph {\bibinfo {title} {Nonlinear differential
  equations and dynamical systems}}}\ (\bibinfo  {publisher} {Springer Science
  \& Business Media},\ \bibinfo {year} {2006})\BibitemShut {NoStop}%
\bibitem [{\citenamefont {Guckenheimer}\ and\ \citenamefont
  {Holmes}(2013)}]{GH13}%
  \BibitemOpen
  \bibfield  {author} {\bibinfo {author} {\bibfnamefont {J.}~\bibnamefont
  {Guckenheimer}}\ and\ \bibinfo {author} {\bibfnamefont {P.}~\bibnamefont
  {Holmes}},\ }\href@noop {} {\emph {\bibinfo {title} {Nonlinear oscillations,
  dynamical systems, and bifurcations of vector fields}}},\ Vol.~\bibinfo
  {volume} {42}\ (\bibinfo  {publisher} {Springer Science \& Business Media},\
  \bibinfo {year} {2013})\BibitemShut {NoStop}%
\bibitem [{\citenamefont {Cressman}\ and\ \citenamefont
  {Vickers}(1997)}]{CV97}%
  \BibitemOpen
  \bibfield  {author} {\bibinfo {author} {\bibfnamefont {R.}~\bibnamefont
  {Cressman}}\ and\ \bibinfo {author} {\bibfnamefont {G.}~\bibnamefont
  {Vickers}},\ }\bibfield  {title} {\bibinfo {title} {Spatial and density
  effects in evolutionary game theory},\ }\href@noop {} {\bibfield  {journal}
  {\bibinfo  {journal} {Journal of theoretical biology}\ }\textbf {\bibinfo
  {volume} {184}},\ \bibinfo {pages} {359} (\bibinfo {year}
  {1997})}\BibitemShut {NoStop}%
\bibitem [{\citenamefont {Vickers}(1989)}]{V89}%
  \BibitemOpen
  \bibfield  {author} {\bibinfo {author} {\bibfnamefont {G.}~\bibnamefont
  {Vickers}},\ }\bibfield  {title} {\bibinfo {title} {Spatial patterns and
  {ESS}'s},\ }\href
  {https://doi.org/http://dx.doi.org/10.1016/S0022-5193(89)80033-5} {\bibfield
  {journal} {\bibinfo  {journal} {Journal of Theoretical Biology}\ }\textbf
  {\bibinfo {volume} {140}},\ \bibinfo {pages} {129} (\bibinfo {year}
  {1989})}\BibitemShut {NoStop}%
\bibitem [{\citenamefont {Vickers}(1991)}]{V91}%
  \BibitemOpen
  \bibfield  {author} {\bibinfo {author} {\bibfnamefont {G.}~\bibnamefont
  {Vickers}},\ }\bibfield  {title} {\bibinfo {title} {Spatial patterns and
  travelling waves in population genetics},\ }\href@noop {} {\bibfield
  {journal} {\bibinfo  {journal} {Journal of Theoretical Biology}\ }\textbf
  {\bibinfo {volume} {150}},\ \bibinfo {pages} {329} (\bibinfo {year}
  {1991})}\BibitemShut {NoStop}%
\bibitem [{\citenamefont {deForest}\ and\ \citenamefont
  {Belmonte}(2013)}]{dB13}%
  \BibitemOpen
  \bibfield  {author} {\bibinfo {author} {\bibfnamefont {R.}~\bibnamefont
  {deForest}}\ and\ \bibinfo {author} {\bibfnamefont {A.}~\bibnamefont
  {Belmonte}},\ }\bibfield  {title} {\bibinfo {title} {Spatial pattern dynamics
  due to the fitness gradient flux in evolutionary games},\ }\href@noop {}
  {\bibfield  {journal} {\bibinfo  {journal} {Physical Review E}\ }\textbf
  {\bibinfo {volume} {87}} (\bibinfo {year} {2013})}\BibitemShut {NoStop}%
\end{thebibliography}%

\end{document}